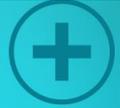

# ETUDE SUR LES PORTAILS ET AGREGATEURS DES RESSOURCES PEDAGOGIQUES UNIVERSITAIRES FRANCOPHONES EN ACCES LIBRE

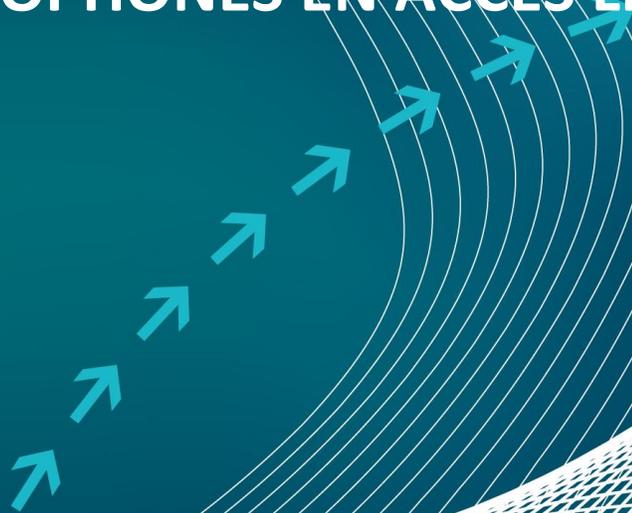

**Mokhtar BEN HENDA**
Septembre 2015

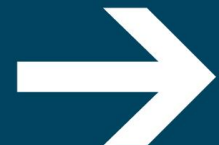

**IDNEUF**
Initiative pour le développement du numérique dans l'espace universitaire francophone

# Table des matières

# 1 RAPPEL DU CONTEXTE

La première réunion des Ministres francophones de l'enseignement supérieur, tenue le 5 juin 2015 à Paris, a permis de discuter de l'état et des perspectives de développement numérique de l'espace universitaire francophone (DNEUF). Les 32 ministres présents ont adopté une déclaration commune[1] appelant entre autres à ce que « les ministres présents ou représentés soutiennent la proposition que les pays et les établissements qui le souhaitent construisent dans les meilleurs délais un portail francophone commun accueillant les ressources numériques universitaires disponibles dans les pays et les établissements intéressés ».

Ce sommet DNEUF a permis de dresser l'état des lieux des ressources numériques dans l'espace universitaire francophone, d'évoquer les besoins et le rôle des universités et de réfléchir à une meilleure mutualisation des ressources existantes.

Dans son alinéa premier, la déclaration commune attribue à l'Agence Universitaire de la Francophonie la tâche d'« animer un groupe d'experts des pays concernés afin d'établir le cahier des charges d'un tel outil [portail francophone commun] et d'en proposer les modes possibles de réalisation ».

Dans cette perspective, une équipe d'experts coordonnée par l'AUF a été constituée pour concrétiser les directives de la déclaration commune et préparer un compte rendu et une démonstration du produit aux Ministres francophones de l'enseignement supérieur qui seront de nouveau réunis en juin 2016 à Bamako.

Dans sa feuille de route, l'AUF a programmé une première réunion de suivi tenue à Paris le 17 juillet 2015 pendant laquelle une série de mesures a été entreprise en vue de préparer le cahier des charges du portail prévu. Parmi ces mesures, dont les aboutissements devraient être présentés à la prochaine réunion de l'équipe d'experts vers la fin de septembre 2015. Parmi ces mesures, il a été convenu de fournir deux types de produits :

1. Une étude exploratoire sur les portails des ressources pédagogiques existants. Ces portails devraient être essentiellement francophones et en accès libre. Quelques portails internationaux seraient toutefois utiles à explorer pour servir de repères de comparaison et de positionnement de l'expérience franco-francophone dans le contexte technologique international des portails de ressources pédagogiques ;

2. Une étude prospective argumentée qui devrait permettre d'établir un choix entre deux solutions technologiques envisagées pour la création du portail francophone commun :

    a) la première solution (jugée plus rapide) consiste à étendre à l'ère francophone les capacités du moteur de recherche développé dans le cadre du portail du numérique dans l'enseignement supérieur (sup-numerique.gouv.fr) ;

    b) la deuxième (jugée possible mais plus couteuse et moins rapide) consiste à créer un portail fédérateur (un portail des portails) qui référence et lie entre eux les principaux sites

---

[1] Déclaration commune des Ministres francophones de l'Enseignement supérieur pour le développement numérique de l'espace universitaire francophone.

universitaires francophones fournissant des ressources pédagogiques ou renvoyant vers les moteurs de recherche qui indexent leurs ressources respectives.

La question de la compilation, du référencement, de l'indexation et de l'accès transparent aux ressources pédagogiques en ligne est donc au cœur du processus de création du futur portail francophone commun. La solution à retenir devrait tenir compte des avantages et des inconvénients de l'une et de l'autre autant sur le plan technique qu'économique et chronologique.

# 2 LIMITES DE L'ETUDE

Le présent rapport répond à la première mesure entreprise le 17 juillet 2015, celle de l'analyse exploratoire des portails et agrégateurs existants des ressources pédagogiques universitaires francophones en accès libre. L'idée en est de fournir un aperçu sur les tendances et les pratiques les plus courantes dans la constitution et l'organisation de portails numériques des ressources pédagogiques en ligne. L'étude de ces tendances aiderait à définir les choix et les conditions idoines pour concevoir le futur portail francophone commun et optimiser ses services de conservation, échange, intégration et mutualisation des ressources pédagogiques dans le cadre technologique distribué des universités francophones. Ce cadre devrait, en conséquence, être interconnecté, transparent et interopérable, traduisant à la fois les spécificités linguistiques et culturelles des institutions partenaires et leurs ambitions de développements technologiques et économiques.

L'élaboration de cette première étude exploratoire des portails tiendrait compte des deux solutions technologiques discutées pendant la réunion du 17 juillet 2015.

1. D'abord, l'alternative d'étendre aux institutions universitaires francophones les capacités du moteur de recherche de France Université Numérique imposera de tenir compte en priorité de l'expérience des UNT (Université Numériques Thématiques) comme acteurs essentiels du portail du numérique dans l'enseignement supérieur (sup-numerique.gouv.fr). Ainsi, la présente étude devrait d'abord viser les portails des UNT comme modèles reproductibles ou extensibles au contexte francophone en analysant leurs choix technologiques, leurs modes d'organisation et leurs modalités d'usage et de communication ;

2. Ensuite, l'étude ne se limitera pas à l'analyse des portails des UNT. L'hypothèse proposée de créer un portail commun des portails francophones à partir des portails universitaires existants, nécessite d'explorer aussi cette éventualité et de proposer une analyse de portails existant qui fonctionnent selon d'autres modèles d'organisation.

A ces deux délimiteurs, deux autres critères guideront le choix des portails à étudier dans ce rapport :

1. D'abord, pour l'AUF, seules les ressources pédagogiques (le terme sera défini plus loin) sont à référencer et non les parcours de formation du type FOAD ou MOOCs. Ceci réduira notre champ d'analyse qui ne tiendra pas compte des portails spécialisés dans ce genre de matériaux didactiques et pédagogiques ;

2. Ensuite, le projet de portail francophone commun sera centré (du moins en première phase) sur le contexte universitaire. Il répondra en priorité à une déclaration des ministres francophones chargés de l'enseignement supérieur et non de ministres de tout autre secteur d'enseignement comme le secondaire ou le primaire. Il est donc attendu que l'étude se limite à l'enseignement supérieur ou du moins à bien distinguer l'enseignement supérieur de l'enseignement de base en cas de portails génériques.

# 3  CHOIX STRATEGIQUES

Concevoir et élaborer un portail de ressources pédagogiques est une tâche laborieuse qui nécessite non seulement un choix approprié des ressources et des services à fournir à la communauté d'utilisateurs cibles, mais aussi une analyse prospective des besoins et des formes d'évolution qui produiraient de la valeur ajoutée pour l'enseignement et la recherche.

Les principaux objectifs d'une étude d'évaluation est d'identifier et de recommander une solution pour un système de portail unifié répondant aux spécifications suivantes : une plateforme intégrée qui fournirait en toute sécurité aux utilisateurs un point central d'accès, de personnalisation et de configuration de l'information et des applications appropriées à leurs rôles au sein de l'université. Le portail offrirait également des moyens normalisés d'agrégation de données. Il permettrait également aux développeurs et aux fournisseurs de services de fournir des s applications et des informations via une variété de plates-formes fixes et mobiles.

La décision de monter un portail francophone de ressources pédagogiques dans le cadre de la commande DNEUF gagnerait à intégrer les trois objectifs complémentaires suivants :

1. ***Objectif économique*** : valoriser le patrimoine numérique local de chaque institution partenaire du projet et construire un réseau de portails interuniversitaires pour réduire les coûts de reproduction et démultiplier les usages des ressources numériques. Cet objectif participera de la création d'une archive institutionnelle vivante et unique regroupant tous les documents qui sont à la fois utilisés quotidiennement et suffisamment importants pour être conservés et réutilisés sur le long terme ;

2. ***Objectif scientifique*** : construire un réseau de portails interuniversitaires qui vise à échanger et à partager les patrimoines numériques de chaque établissement avec des partenaires locaux ou internationaux globalement ou sur un thème particulier, gratuitement ou en fonction de droits d'accès. Le portail commun des ressources pédagogiques contribuerait ainsi à enrichir l'activité de la recherche et de la publication scientifiques ;

3. ***Objectif pédagogique*** : intégrer le portail commun des ressources pédagogiques aux cursus universitaires et aux programmes de formation en ligne et le rendre accessible aux enseignants et aux étudiants à travers des profils d'accès et des modules d'enseignement. Le portail ne sera pas ainsi un simple réservoir d'objets pédagogiques, dissocié de tout contexte d'usage académique réel et concret.

A ces trois grands objectifs stratégiques, deux autres objectifs (ou exigences) d'ordre technologique et culturel sont également à prévoir dans la construction d'un portail francophone commun de ressources pédagogiques gratuites :

1. D'abord, une exigence technique (et technologique) de convergence et de cohérence avec les pratiques internationales dans la conception et la diffusion des ressources pédagogiques numériques en ligne. Inspirés des grandes vagues d'innovations technologiques matérialisées à travers les centres de données, les bibliothèques numériques, les archives ouvertes, les réservoirs ou dépôts d'objets pédagogiques, etc., les portails des ressources pédagogiques deviennent progressivement une catégorie de services à part entière qui héritent des mêmes principes d'intégration, de convergence et d'interopérabilité qu'on trouve dans les autres catégories de systèmes d'information numériques en réseau. En tant qu'objets numériques, toutes les ressources pédagogiques sont concernées par les techniques de description, d'organisation, de traitement et d'exploitation définies par les normes, standards et spécifications internationales. Un portail de ressources pédagogiques francophones gagnerait à tenir compte des référentiels normatifs internationaux dans la conception de son modèle opératoire, de son mode de gouvernance et de ses processus de traitement, d'analyse et d'exploitation des données ;

2. Ensuite, une exigence de cohérence et de concordance entre les acteurs/bénéficiaires du portail. Les concepteurs du portail commun de ressources pédagogiques francophones devraient faire preuve d'une compréhension des réalités profondes des acteurs/bénéficiaires et d'une projection de leurs besoins non seulement d'un point de vue éducatif, mais aussi culturel, linguistique et économique. La Francophonie est un champ géostratégique vaste et composite. Un portail de ressources pédagogiques francophones devrait permettre de gérer mieux que quiconque la complexité de la diversité. Il devrait servir de foyer de convergence à toutes les tendances et les exigences exprimées par les individus et les communautés de pratiques scientifiques en Francophonie. Le patrimoine des ressources pédagogiques existantes, à la fois développées et échangées via les systèmes et les plates-formes pédagogiques francophones (anciennes et courantes) et les services numériques locaux des universités francophones doivent être capitalisées pour l'enseignement, la formation et l'apprentissage (cours en lignes, réservoirs d'objets pédagogiques, bases de données de REL, scénarios pédagogiques, etc.). Ces matériaux pédagogiques et scientifiques fourniront une matière expérimentale de haute importance pour la mise en œuvre du portail et son acceptabilité par les populations cibles.

# 4 POUR UN PORTAIL COMMUN DE RESSOURCES PEDAGOGIQUES FRANCOPHONES

Selon la définition du *Joint Information Systems Committee* (JISC), un portail est « un ''service réseau'' qui rassemble une offre de contenu provenant de plusieurs sources distinctes et met en œuvre des technologies telles que la recherche fédérée, le moissonnage d'informations, les services d'alerte, et compile les résultats dans un format unifié afin de les présenter à l'usager. Cet affichage se fait habituellement via un navigateur Web, bien que d'autres moyens soient également possibles. Pour les usagers, un portail est un point d'accès, éventuellement personnalisable, où une recherche peut être menée sur plus d'une ou plusieurs sources d'information, où les résultats affichés sont fusionnés et uniformisés. L'information peut y être également présentée par d'autres moyens comme par exemple des services d'alerte, des listes de discussion ou des liens vers des documents électroniques ou des supports pédagogiques ».

Dans le livre blanc de la société e-Zest intitulé « *Enterprise Portal Evaluation & Implementation* », il est spécifié que « Un portail […] est un cadre d'intégration d'information, de personnes et de processus dans des frontières organisationnelles. Le portail fournit un point d'accès unifié et sécurisé, souvent sous la forme d'une interface utilisateur sur le Web, conçu pour agréger et personnaliser l'information grâce à des portlets[2] spécifiques à l'application. L'une des marques de fabrique des portails d'entreprise est la contribution et la gestion décentralisée du contenu qui garantit que l'information soit à jour en permanence. Une solution réussie de portail d'entreprise peut agréger des contenus et des applications de façon dynamique en vue de fournir différents types de services proposant un libre accès à l'information […] Les utilisateurs finaux peuvent personnaliser un portail d'entreprise en ajoutant et en soustrayant des sources et des applications d'information internes et externes. En théorie, la personnalisation fournit aux employés toutes les informations dont ils ont besoin pour faire leur travail plus rapidement et plus efficacement que s'ils devraient rechercher les données et les applications par eux-mêmes.

Les portails en général peuvent être destinés à plusieurs domaines de connaissance ou à divers groupes d'intérêt. On peut trouver des portails d'entreprise, des portails de divertissement, des portails de jeux, des portails politiques, des portails éducatifs, etc. On peut même classer les portails selon qu'ils soient généraux, verticaux ou spécialisés.

Les modes d'organisation des portails varient également selon les cas. Certains s'établissent comme des catalogues de ressources organisées en silos, intégrées dans des bases de données centralisées et classées par disciplines. D'autres, organisées par niveaux thématiques, renvoient vers des bases de données réparties, etc. Si on s'intéresse au succès des portails, on s'apercevra que la majorité cumule une triple méthode pour faciliter l'accessibilité aux ressources : ils disposent d'une indexation souvent inspirée des schémas et profils d'application des métadonnées normalisés comme Dublin Core, d'un modèle de référencement et d'un moteur de recherche soutenues par des algorithmes de avancés.

Cependant, bien que les portails prennent différentes significations selon la perspective qu'on veut leur donner, ils gardent une caractéristique commune, celle de constituer un endroit où sont cumulées d'énormes ressources d'information (locales ou hébergées sur plusieurs sites répartis). Ces ressources sont soumises à une chaine de référencement et de production qui permet de les décrire, indexer, analyser, classer, conserver et communiquer selon les besoins des usagers.

Dans le domaine de l'éducation, de plus en plus de portails éducatifs sont accessibles en ligne bien que beaucoup d'entre eux ne remplissent pas les conditions qui devraient caractériser un véritable portail éducatif. En plus d'être un dispositif de gestion, d'organisation et d'administration de cursus et de formations universitaires, un véritable portail éducatif devrait être en mesure de fournir un environnement de collaboration pour le développement, l'évaluation et le partage des matériaux scientifiques et didactiques et des ressources de formation et d'apprentissage. Un portail éducatif devrait donner l'accès à une multitude de ressources éducatives et d'information qui ont un potentiel de compléter, de manière efficace, le processus d'enseignement et d'apprentissage. Pour enseigner ou apprendre, l'enseignant comme l'apprenant utilisent des ressources pédagogiques et scientifiques sous forme de supports de cours, de diaporamas, de rapports, d'articles, de conférences, de livres, etc. Un portail éducatif devrait aider à mutualiser ces ressources de façon transparente, librement ou via des profils et des droits d'accès, pour qu'un enseignant qui prépare un cours trouve des exemples et des ressources didactiques, un étudiant trouve des exercices et des compléments d'information à son parcours d'apprentissage, un chercheur trouve des publications utiles à sa recherche scientifique, etc.

---

[2] Un portlet est une application informatique que l'on peut placer dans un portail web pour servir de conteneur. C'est un objet qui affiche un bloc sur une page web. Du point de vue de l'interface Web, le portlet est vu par l'utilisateur comme un composant qu'il peut afficher où il veut dans la vue personnalisée de son portail.

Concrètement, dans le projet DNEUF, chaque université peut construire un portail permettant d'archiver, de diffuser son patrimoine numérique qu'il soit documentaire, scientifique ou pédagogique, conformément à sa politique d'établissement, et avoir accès aux documents des établissements qui utilisent les mêmes protocoles. Or, aujourd'hui, nous assistons de plus en plus à l'émergence de portails communicants qui disposent de modèles d'architecture plus complexes et spécifiques répondant aux attentes des communautés universitaires plus larges comme celles des MOOCs ou des consortiums et regroupements universitaires. Dans ces alliances stratégiques universitaires et ces communautés de pratiques virtuelles et dispersées, la diversité des plates-formes technologiques, des réseaux de transmissions de données, des supports de cours, des formats de fichiers, des outils applicatifs, etc. ne doivent en aucun cas constituer un blocage à la convergence, transparence et interopérabilité des services et des contenus. Les solutions technologiques, avec l'appui de la normalisation internationale, ont largement contribué à rendre entièrement transparentes l'intégration des services, la convergence des systèmes et la fusion des ressources. Les portails communicants en sont une démonstration concrète.

Le concept de « réseau de portails communicants » résulte des innombrables innovations dans les domaines des technologies numériques, des réseaux de télécommunication et du génie logiciel. Le principe de « réseau de portails communicants », proposé dans le cadre universitaire français par l'Université numérique thématique ingénierie et technologie (UNIT), est un exemple bien connu qui répond justement aux attentes de l'intégration transparente des systèmes d'information distribués.

Dans une pareille perspective, la construction d'un réseau de portails communicants permettrait d'améliorer le partage de ressources et faciliterait la création de nouvelles connaissances, tant sur le plan pédagogique que scientifique et culturel :

- Les enseignants peuvent s'appuyer sur les cours, les travaux pratiques, les exercices, les applications, etc. déjà réalisés par leurs collègues pour créer de nouveaux contenus plus riches, plus clairs et plus innovants, tout en évitant de réinventer la roue.

- Les étudiants peuvent découvrir de nouveaux éclairages sur les concepts qu'ils doivent apprendre pour préparer leurs examens à partir d'un plus grand nombre d'exercices, ces derniers ayant été finement et richement indexés par des professionnels compétents.

- Les chercheurs peuvent accéder facilement à l'ensemble de la production scientifique des établissements partenaires, cette production ne se limitant pas aux articles publiés dans les grandes revues, puisqu'elle comprend aussi les actes de colloques, les chapitres dans les publications collectives, les thèses, les mémoires, etc. Ils peuvent ainsi mieux cibler leurs propres recherches ou trouver de nouvelles idées.

- Les établissements peuvent mieux suivre les évolutions dans les domaines qui les intéressent.

C'est cette optique de portail ouvert et connecté sur l'univers académique que le futur portail francophone commun devrait mettre en pratique. Sa construction est de loin moins complexe et couteuse que de créer un réservoir central des ressources numériques détenues par les universités francophones et accessibles via un portail centralisé. Une démarche de convergence entre les serveurs des ressources existantes économiserait des coûts de duplication et de montage de nouveaux dispositifs pédagogiques et ne toucherait pas à l'autonomie des universités partenaires dans leurs modèles de gouvernances locales. Celles-ci ne seraient pas contraintes de changer radicalement leurs politiques de gestion et d'organisation de leurs ressources pédagogiques. Le portail commun serait juste une couche logicielle supérieure constituée d'API et de protocoles d'indexation-recherche qui permettront d'interroger l'ensemble des serveurs des universités partenaires sans se soucier de leurs particularismes technologiques.

C'est justement au niveau de cette couche logicielle d'indexation-recherche qu'un travail de conception-développement devrait avoir lieu pour le développement et la maintenance d'un réseau de portails communicants interuniversitaire. Plusieurs exemples de portails fédératifs répondent à ce critère de répartition de ressources et de centralisation de mécanismes d'indexation-recherche dont notamment SUDOC (http://www.sudoc.abes.fr/), le catalogue du Système Universitaire de Documentation, réalisé par les bibliothèques et centres de documentation de l'enseignement supérieur et de la recherche[3]. Les portails des universités numériques thématiques constituent un autre exemple de portails proposant des ressources distribuées dans lesquels un processus de référencement et d'indexation conforme à des référentiels normatifs est appliqué.

# 5 CHAINE DE REFERENCEMENT ET D'INDEXATION-RECHERCHE

Plusieurs solutions technologiques sont en mesure de fournir ce genre de scénarios, sous réserve que les serveurs des universités assurent, dans le cadre d'une convention cadre, un minimum de cohérence et de convergence dans le choix et l'application des normes et protocoles de traitement de données. Une chaine de référencement et d'indexation-recherche interopérable (et pas nécessairement homogène) permettrait plus facilement à chacun des membres du réseau d'avoir accès aux documents des établissements partenaires qui utilisent les mêmes normes et les mêmes protocoles.

Il existe des moteurs d'indexation-recherche uniquement consacrés aux pages web. Ils ne nous intéressent pas beaucoup dans cette étude plutôt orientée vers l'analyse des moteurs d'indexation de tous types de ressources multimédias. Il existe aussi des moteurs d'indexation-recherche plus génériques qui peuvent indexer n'importe quel type de ressource et administrer pratiquement tous les produits de gestion de contenus et de gestion de documents, qu'ils soient open source ou non.

---

[3] SUDOC comprend plus de 10 millions de notices bibliographiques qui décrivent tous les types de documents (livres, thèses, revues, ressources électroniques, documents audiovisuels, microformes, cartes, partitions, manuscrits et livres anciens...)

Ce qui nous intéresse ici, c'est l'utilisation d'un moteur d'indexation-recherche dans la gestion des données d'une plateforme web. Un tel moteur a de nombreux atouts :

- Il est plus performant qu'un SGBD sur certaines typologies de requêtes complexes ;

- En particulier, il excelle dans des requêtes qui réunissent contenus structurés et contenus non-structurés. Par « contenus non-structurés », on entend les textes et documents ;

- Il supporte de très gros volumes sans dégradation des performances. Typiquement plusieurs dizaines de millions d'items sont monnaie courante ;

- Sa fonction n'est pas de stocker, ni de gérer l'information, il donne juste un moyen de la retrouver par la recherche.

Intégrer un moteur de recherche en complément d'un SGBD est dans la logique courante de ne pas s'appuyer systématiquement sur la même panoplie d'outils standards, mais d'utiliser au contraire le meilleur outil pour chaque fonction. La base de données reste par contre le lieu de référence de gestion de l'information, mais les objets à rechercher sont passés au moteur pour indexation. Ils peuvent être par exemple exportés au format XML, et analysés dans cette forme par l'indexation. Ou bien les APIs d'indexation peuvent être appelées directement par un traitement batch lisant dans la base. Mais on préfèrera nettement considérer la base de données et la fonction d'indexation-recherche comme deux sous-systèmes disjoints, qui doivent interagir uniquement au travers du middleware ou bien par des APIs bien définies, et non un traitement batch « à cheval » sur les deux.

Rappelons que le fonctionnement standard d'un moteur de recherche est communément défini par trois processus successifs et itératifs :

1. **L'exploration** ou *crawling* : un moteur de recherche est d'abord un outil d'indexation, c'est-à-dire qu'il dispose d'une technologie de collecte de documents à distance sur les sites Web, via un outil que l'on appelle robot ou bot ou spider. Un robot d'indexation écume récursivement tous les hyperliens qu'il trouve pour récupérer les ressources considérées intéressantes (selon un algorithme de pondération ou scoring). L'exploration est lancée depuis un point de départ généralement constitué de la page d'annuaire web. Pour le projet de portail francophone commun, le moteur de recherche doit disposer d'un robot capable de parcourir les liens entre les serveurs de données des institutions partenaires pour indexer leurs contenus. Cette possibilité est définie dans la programmation de l'algorithme qui fixe le mode de fonctionnement du robot.

2. **L'indexation** : tous les mots significatifs dans les ressources récupérées sont extraits vers une base de données organisée comme un gigantesque dictionnaire inverse dans un ordre alphabétique qui commence par les symboles et continue par les chiffres puis lettres. Cet index permet de retrouver rapidement l'emplacement des termes significatifs dans leurs documents sources. Les termes non significatifs (mots vides) sont exclus de l'indexation. Les termes significatifs sont associés à un *poids* qui reflète, entre autres, la probabilité d'apparition du mot dans un document, l'importance de sa position, l'actualisation de sa source et le « pouvoir discriminant de ce mot » dans une langue. Pour réaliser cette fonction dans l'envergure d'un portail francophone commun, la technique des réservoirs des métadonnées pédagogiques selon le modèle des archives ouvertes (et le protocole OAI-PMH) est très recommandée. Nous y reviendrons dans les points suivants ;

3. **La recherche** : cette phase correspond à la fonction de formulation d'équations de recherche et de restitution des résultats. En passant par l'index (liste alphabétique des mots significatifs), un algorithme spécifique permet d'identifier les documents répondent le mieux aux mots contenus dans la requête. Le résultat de recherche est alors présenté par ordre de pertinence. « Les algorithmes de recherche font l'objet de très nombreuses investigations scientifiques. Les moteurs de recherche les plus simples se contentent de requêtes booléennes pour comparer les mots d'une requête avec ceux des documents. Mais cette méthode atteint vite ses limites sur des corpus volumineux. Les moteurs plus évolués sont basés sur le paradigme du modèle vectoriel. Ils utilisent la formule TF-IDF (*Term Frequency-Inverse Document Frequency*)[4] pour mettre en relation le poids des mots dans une requête avec ceux contenus dans les documents. Cette formule est utilisée pour construire des vecteurs de mots, comparés dans un espace vectoriel, par une similarité cosinus. Pour améliorer encore les performances d'un moteur, il existe de nombreuses techniques, la plus connue étant celle du *PageRank* de Google qui permet de pondérer une mesure de cosinus en utilisant un indice de notoriété de pages. Les recherches les plus récentes utilisent la méthode dites d'analyse sémantique latente qui tente d'introduire l'idée de cooccurrences dans la recherche de résultats (le terme « voiture » est automatiquement associé à ses mots proches tels que « garage » ou un nom de marque dans le critère de recherche).

L'évolution continue des systèmes d'information (analogiques ou numériques) a donné lieu à des techniques d'indexation-recherche qui évoluent au rythme de la complexité des modes d'organisation des fonds et des collections des données. Plus les ressources sont distribuées sur les réseaux, plus les techniques de localisation, référencement, indexation et recherche gagnent en complexité et mais aussi en performance.

Parmi les solutions qui ont marqué l'histoire des systèmes d'information distribué, la norme Z39.50 a sans doute été l'une des plus utilisées jusqu'aux années 2000, quand le protocole OAI-PMH (*Open Archive Initiative-Protocol for Metadata Harvesting*) a fait son apparition. Rappelons dans ce sens que le protocole Z39.50 a été développé vers la fin des années 1980 par le Zig (Z39-50 Implementers Group) et gérée par la ''Z39-50 maintenance Agency'' hébergée par la Library of Congress. Devenu une norme ISO en mars 1997 (ISO 23950), la Z39.50 définit un système client/serveur basé sur un service et un protocole pour la recherche et le transfert d'informations en formats MARC structurés.

Le protocole Z39.50 spécifie les procédures et les formats pour permettre à un client de requêter une base de données proposée par un serveur, d'identifier les informations correspondant aux critères de la recherche et de récupérer les informations identifiées. Parmi ses caractéristiques c'est qu'à la différence d'un moissonneur comme OAI-PMH, il fonctionne en mode synchrone, c.-à-d. les requêtes émanant du client sont exécutées en direct sur le serveur distant et les résultats sont rapatriés instantanément en retour. Cette configuration n'est plus d'actualité pour gérer des masses considérables d'information et des trafics gigantesques de données issus des transactions de recherche au niveau des serveurs.

---

[4] Le TF-IDF est une méthode de pondération (mesure statistique) utilisée en recherche d'information et en particulier dans la fouille de textes. Elle permet d'évaluer l'importance d'un terme contenu dans un document, relativement à une collection ou un corpus. Le poids augmente proportionnellement au nombre d'occurrences du mot dans le document. Il varie également en fonction de la fréquence du mot dans le corpus. Des variantes de la formule originale sont souvent utilisées dans des moteurs de recherche pour apprécier la pertinence d'un document en fonction des critères de recherche de l'utilisateur.

Au fil du temps, de nouvelles solutions plus performantes et compétitives se font distinguer sur le marché des technologies numériques. Le protocole OAI-PMH émerge du lot grâce à un large consensus sur sa capacité à partager des ressources numériques au sein de plusieurs variantes de systèmes d'information distribuée comme les archives ouvertes, les portails numériques, les centres de données, les réservoirs d'objets pédagogiques, les dispositifs d'enseignement à distance, les sites des réseaux sociaux, les systèmes d'édition de contenus (CMS), etc. Sa force majeure est sans doute celle de pouvoir séparer les ressources de leurs données de description pour créer et gérer des quantités importantes de métadonnées (réservoirs de métadonnées liées à des ressources distribuées) grâce à un protocole de moissonnage (indexation) sur réseaux.

# 6 OAI-PMH : UN PROTOCOLE DE PARTAGE DE RESSOURCES DISTRIBUEES

OAI-PMH, est un protocole international pour le moissonnage des métadonnées des archives ouvertes. Elaboré par l'*Open Archive Initiative* en 1999, il permet de créer, d'alimenter et de tenir à jour des réservoirs d'enregistrements qui signalent, décrivent et rendent accessibles des documents sans les dupliquer ni modifier leur localisation d'origine.

OAI-PMH est largement implémenté dans de nombreux systèmes intégrés de gestion de bibliothèque, de gestion électronique de documents, de création de bibliothèque virtuelle, etc. S'appuyant sur des standards existants (HTTP, XML, Dublin Core, etc.), il permet de rapprocher et de faire communiquer entre elles des bases de données hétérogènes.

Le protocole OAI-PMH est un enjeu essentiel sur les questions d'interopérabilité et de diffusion. Il permet à chaque établissement de conserver ses documents sur ses propres serveurs tout en mettant à disposition les métadonnées et les informations d'accès. Il accélère les recherches des ressources hétérogènes et dispersées sans besoin de les regrouper ni de les dupliquer. Chaque établissement conserve ainsi ses seuls documents sur ses serveurs, dans leur dernière version indépendamment des applications utilisées pour leur gestion.

Avec ce protocole, les métadonnées sont mises à disposition sur un serveur (elles sont publiées). Elles peuvent être agrégées avec d'autres et interrogées par des moteurs de recherche spécialisés ou généralistes. La conformité au protocole OAI-PMH permet aux différents portails de devenir communicants : ils rendent publiques leurs métadonnées et peuvent eux-mêmes valoriser les métadonnées d'autres portails. En raison de sa souplesse, son champ d'utilisation s'étend maintenant bien au-delà des seuls contenus scientifiques. Il devient un modèle de référence pour plusieurs plates-formes et logiciels open source qui permettent de capturer et de décrire des contenus numériques : Archimede (Bibliothèque de Laval), Fedora (Université de Cornell), Dspace (Bibliothèques du MIT), ePrints (Université de Southhampton), ORI-OAI (UNT), etc.

## 6.1 MODELES D'EXPLOITATION D'OAI-PMH

Il existe de nombreuses solutions *open source* conformes au protocole d'échange OAI-PMH qui offrent un ensemble de services permettant de gérer et de diffuser des ressources électroniques distribuées. Eprints, DSpace, Fedora, ORI-OAI, etc. sont autant d'exemples de systèmes ouverts qui, malgré leur différences fonctionnelles, proposent des services orientés vers la gestion de collections de ressources numériques distribuées.

Nous faisons ci-après un bref comparatif entre trois des solutions les plus citées dans la littérature scientifique, à savoir Eprints, Dspace et ORI-OAI, pour en dégager les points forts et les points faibles capables d'orienter les choix dans la construction du portail commun des ressources pédagogiques francophones. Un tableau synthétique résumera ensuite les points de comparaisons entre les trois outils en question.

### 6.1.1 EPRINTS

Crée en 2000 à l'Université de Southampton, le logiciel Eprints a été le premier logiciel d'auto archivage. Crée à l' origine dans le but d'archiver des articles scientifiques, il permet aujourd'hui de soumettre, de consulter et de gérer des documents électroniques.

Assurant la compatibilité et l'interopérabilité avec toutes les archives ouvertes respectant le protocole OAI-PMH, le logiciel Eprints offre de nombreuses fonctionnalités. Il permet entre autres de traiter plusieurs types de documents (thèses, ressources pédagogiques, etc.), d'établir le modèle de métadonnées en fonction du type de publication et d'organiser les documents dans des ensembles de thématiques.

Le logiciel offre également des options de recherche de documents selon plusieurs critères. Offrant des interfaces conviviales disponibles en français, Eprints est facile à prendre en main et s'adapte bien aux besoins des institutions. De plus, sa mise en place et sa maintenance ne requièrent pas l'intervention d'un grand nombre de main d'œuvre.

Cependant, Eprints n'offre pas de solutions de pérennisation des documents. Le fait qu'il ne gère pas les changements de formats, certains documents peuvent s'avérer illisibles à plus long terme. La version standard basique nécessite de nombreux développements en PERL pour pallier à ces lacunes.

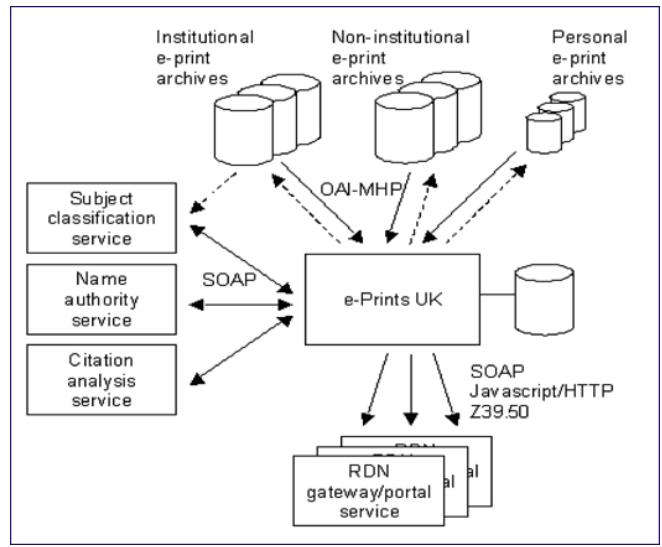

Source : ePrints.UK (http://www.ariadne.ac.uk/issue35/martin)

## 6.1.2 DSPACE

Dspace est un logiciel libre développé en 2002 à l'initiative du MIT et des laboratoires HP à Cambridge. Il permet la construction d'archives et de bibliothèques électroniques. Fréquemment utilisé par les universités ou les organismes de recherche pour stocker des collections d'articles, de thèses et de livres en accès libre ou restreint, Dspace peut aussi archiver des photos, des enregistrements sonores ou des vidéos.

Le logiciel permet également de gérer des communautés et des sous-communautés d'utilisateurs, de gérer des collections de documents électroniques et de générer différents workflows de publication comprenant l'auto-archivage.

DSpace s'affirme compatible avec le modèle de référence OAIS pour l'archivage pérenne des données numériques grâce à des fonctionnalités permettant la vérification de l'intégrité des fichiers entreposés. Les métadonnées des ressources sous Dspace sont exposées selon le protocole OAI-PMH, aux formats Dublin Core ou METS, et selon le format OpenURL COinS (une convention pour englober les métadonnées dans le HTML). Le format d'échange des données et métadonnées OAI-ORE est également supporté.

Nonobstant, Dspace présente quelques contraintes techniques. D'un point de vue fonctionnel, il ne prend pas en compte l'obsolescence des formats de documents, ce qui pourrait poser un problème si le format du document venait à disparaître ou que le logiciel pour le lire ne soit plus édité. DSpace dépend d'une série d'autres projets Java, par exemple, Cocoon, chose qui rendrait sa gestion plus complexe. Tout en étant puissant, Dspace reste relativement difficile à maîtriser, à mettre en œuvre et à maintenir. Différentes sociétés de service spécialisées et des développeurs en *freelance* offrent leurs services pour seconder les services informatiques.

D'un point de vue ergonomique, de nombreuses formes de personnalisations de l'interface utilisateur ne sont pas possibles sans des compétences techniques avancées. Dspace ne permet pas par exemple à l'utilisateur de modifier un champ de données dans le processus de soumission, une fois le champ a été validé. Dans ce cas de figure, l'utilisateur doit supprimer complètement l'entrée par exemple un mot-clé, un titre, un auteur, etc., afin de modifier le champ.

Du point de vue métadonnées, DSpace ne gère que Dublin Core, et ne permet pas de prendre en compte d'autres schémas de métadonnées. Dans le projet de la constitution d'un portail commun de ressources pédagogiques francophones, ce point est stratégique. La norme Dublin Core permet effectivement de décrire une typologie variée de documents mais dans certains cas cette description n'est pas assez précise. Notamment dans le cas d'une ressource pédagogique, l'utilisation de normes plus détaillées, comme LOM ou MLR est plus adaptée et plus concise.

La communauté DSpace est organisée selon un modèle voisin de celui de la fondation Apache. Elle repose sur une communauté d'utilisateurs. Les développements sont faits par différents contributeurs, et ajoutés à la distribution après contrôle par une équipe de valideurs.

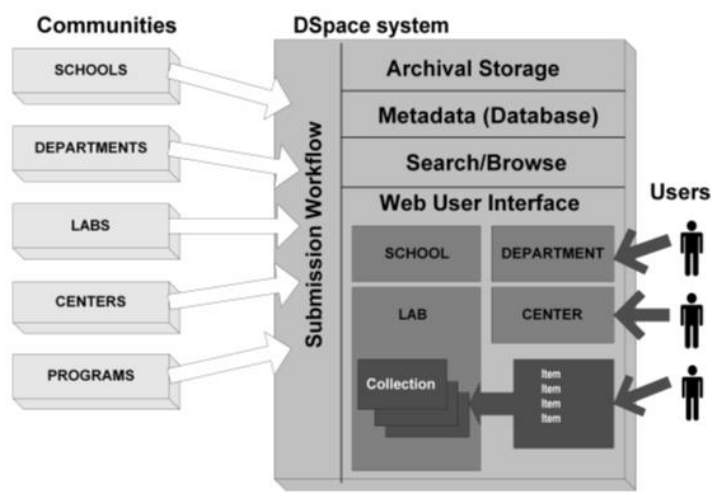

## 6.1.3 ORI-OAI

Le logiciel ORI-OAI est un outil de référencement et d'indexation pour un réseau de portails OAI. Issu d'un projet inter-UNT (Universités Numériques Thématiques), ORI-OAI a été développée par quatre établissements universitaires français : l'INP de Toulouse, l'INSA de Lyon, l'Université de Valenciennes et l'Université de Rennes 1.

Cette initiative française émane du dilemme ressenti par les UNT (Universités Numériques Thématiques) et les UNR (Universités Numériques en Région) face à la question du partage des ressources numériques dans des systèmes interopérables, accessibles depuis les ENT (Environnements numériques de travail) des établissements universitaires et de recherche.

L'idée *princeps* de cette initiative était la mise en place d'un système ouvert capable de :

- gérer tous les documents numériques produits par les établissements universitaires,

- les partager avec d'autres établissements,

- les valoriser par une indexation professionnelle,

- les rendre accessibles, à distance et selon les droits définis, dans des interfaces ergonomiques.

Sur le site docinsa.insa-lyon.fr/ori/, les points-clés de ce système sont résumés comme suit :

- **Référentiel unique**, pour les ressources numériques pédagogiques, scientifiques et documentaires de l'établissement (ressources documentaires acquises ou produites par l'établissement). Le référentiel ne se substitue pas aux diverses plates-formes qui peuvent utiliser et publier ces mêmes ressources (plates-formes pédagogiques, sites web des laboratoires …), mais il gère la forme canonique de la ressource numérique,

- **Système de recherche avancée**, multicritère (métadonnées, texte intégral),

- **Accès thématiques aux ressources**, selon des classifications simplifiées qui exploitent la classification Dewey,

- Système de gestion et de publication des ressources numériques

    - Publication web avec gestion des droits d'accès,

    - Description des ressources selon les normes AFNOR LOM-FR, TEF, DC …, en relation avec les autres systèmes documentaires pour le partage des tables d'autorité (SUDOC, STAR …),

    - Indexation selon les classifications usitées dans les bibliothèques universitaires (Dewey …) et exploitées par les classifications spécifiques des UNT,

    - Archivage des ressources numériques,

- **Système de production** impliquant les acteurs concernés, dans des procédures élaborées matérialisées par des workflows ; gestion des versions et accès aux versions natives des documents pour leurs créateurs,

- **Système de partage**, fondé sur l'échange de métadonnées selon le protocole OAI-PMH, permettant de fonctionner au sein d'une communauté constituée (UNT par exemple) en réseau de portails,

- **Système open-source**, libre de droit, documenté et installable simplement.

En termes techniques, le logiciel ORI-OAI possède une architecture modulaire constituée de huit modules indépendants, interconnectables au travers de Web Services. Ces modules sont :

- *ORI-OAI-workflow* pour la gestion du workflow de saisie

- *ORI-OAI-md-editor* pour l'interface de saisie des métadonnées

- *ORI-OAI-indexing* pour l'indexation des ressources

- *ORI-OAI-search* pour la recherche de documents locaux et distants

- *ORI-OAI-vocabulary* qui gère les différentes classifications et vocabulaires

- *ORI-OAI-repository* pour exposer les fiches de métadonnées via le protocole OAI-PMH

- *ORI-OAI-harvester* pour la moisson d'autres entrepôts via le protocole OAI-PMH

- *ORI-OAI-nuxeo* solution de stockage de documents dans ORI-OAI

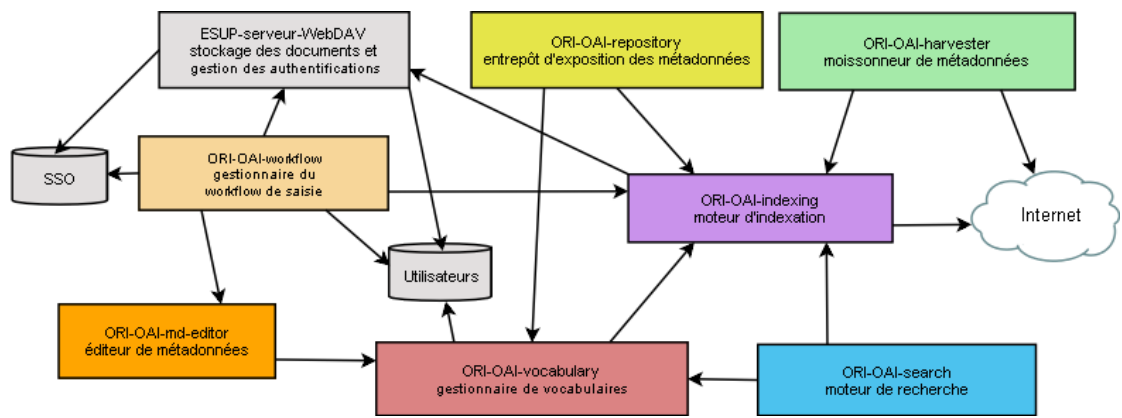

Source : ORI-OAI (wiki.ori-oai.org)

Chacun de ces modules a un rôle bien défini dans le système et communique avec les autres au travers de Web Services exposés en frontal de chaque composant. Ce choix permet une grande souplesse dans d'éventuelles déclinaisons d'architectures et de langages de programmation. En effet, cette architecture technique a été notamment pensée pour permettre à des logiciels extérieurs de dialoguer avec différents éléments du système sans nécessiter l'installation de tous les modules d'ORI-OAI.

Cette configuration propose d'agréger des métadonnées en provenance d'autres établissements, de les exposer pour qu'elles puissent être collectées et de les diffuser au travers d'un portail de recherche. Les fonctionnalités d'agrégation et d'exposition sont, respectivement, gérées au travers des modules ORI-OAI-Harvester et ORI-OAI-Repository qui utilisent le protocole OAI-PMH. Cette configuration est adaptée aux besoins des UNT ou des UNR qui souhaitent proposer une application pour l'ensemble de leurs établissements membres.

Rappelons que l'intégration d'ORI-OAI dans un établissement quelconque s'exprime par la complémentarité avec les applications existantes. Il ne s'agit pas de remplacer divers outils par un autre, mais d'utiliser l'existant pour faire fonctionner un nouvel outil. Par exemple :

- les référentiels de l'établissement (tel que le LDAP…) sont utilisés par ORI-OAI ;

- de nouveaux référentiels peuvent voir le jour et il faut les intégrer au système d'information pour qu'ils puissent être utilisés dans d'autres applications ;

- ORI-OAI est intégrable dans l'Environnement Numérique de Travail (ENT) de l'établissement, ce qui évite de désorienter les utilisateurs, qui ne seront de ce fait pas confrontés à une nouvelle interface de travail ;

- ORI-OAI est interopérable avec l'ensemble des briques du système d'information de l'établissement (telles que Moodle, GRAAL, HAL, les applications de la scolarité, etc…) ;

- ORI-OAI peut devenir le référentiel des documents numériques du SI (plateforme pédagogique, moteur documentaire, sites web de communication…) quand toutes les fonctions de conservation, sur le long terme, sont garanties.

## 6.1.4 COMPARATIF DES PRINCIPAUX OUTILS CONFORMES OAI-PMH

|  | **DSpace** | **EPrints** | **ORI-OAI** |
|---|---|---|---|
| Informations générales | | | |
| **Fournisseur** | DuraSpace, association à but non-lucratif de droit américain | School of Electronics and Computer Science de l'Université de Southampton (Royaume-Uni) | Consortium français d'établissements d'enseignement supérieur et de recherche |
| **Date de création** | 2002 (projet commun entre le MIT et HP) | 2001 | 2006 (fusion de plusieurs projets) |
| **Langues** | Anglais, français, Autres | Anglais, autres langues par addons de contributeurs | Français, anglais |
| **Licence** | BSD | GPL v2 | GPL v3 |
| Informations documentaires | | | |
| **Normes supportées** | Dublin Core | - Dublin Core<br><br>- Autres normes | - Dublin Core, LOM, LOMFr, SupLomFr, TEF avec composants METS, CDM<br><br>- Dewey, CDU, Rameau, MeSH avec gestion d'un format pivot |
| **Ressources gérées par défaut** | Publications scientifiques, thèses, autres documents | Publications scientifiques, autres documents (nombreux types de contenus par défaut) | Ressources pédagogiques, thèses, publications scientifiques (avec Dublin Core étendu), catalogue de formation, autres documents |
| **Edition des données** | Formulaire de soumission configurable ;<br><br>Gestion des listes d'autorité par plugin | Formulaires personnalisables ;<br><br>Gestion des listes d'autorité ;<br><br>Autocomplétion ; Lien avec ROMEO | - Plusieurs formulaires fournis par défaut selon les ressouces gérées ;<br><br>- Possibilité de personnaliser les formulaires d'édition par l'éditeur XML intégré ;<br><br>- Gestion des listes d'autorité |
| **Recherche** | Recherche et filtrage construit autour du moteur libre Lucene ; Recherche avancée sur les champs ; | Recherche globale ; Filtrage ; Recherche avancée ; Recherche | Recherche et filtrage construit autour du moteur libre Lucene et SOLR ; Recherche avancée sur les |

|  | Possibilité de rebonds | thématique | champs ; Recherche à facettes (v2.0) |
|---|---|---|---|
| **Stockage et arcihvage** | - Répertoires de fichiers ;<br>- Système de fichiers distribué | Répertoire de fichiers | - Répertoire de fichiers ;<br>- Base documentaire CMIS (norme des outils de Gestion électronique de documents) |
| **Référencement web** | Lien possible avec Google Scholar | Google Scholar intégré | Intégré en tant qu'entrepôt OAI-PMH |
| Informations techniques ||||
| **Plateforme** | Unix, Linux, Windows, Mac | Unix, Linux, Windows | Unix, Linux, Windows |
| **Installation** | Fichier unique | Fichier unique | - Fichier unique pour l'installation express ;<br>- Installation avancée et modulaire par système de gestion des versions CVS ;<br>- Possibilité de placer chaque module sur un serveur et outil de gestion de base de données différent |
| **Base de données** | MySQL, Oracle, PostGreSQL | MySQL, PostGreSQL | MySQL, PostGreSQL |
| **Liens avec d'autres logiciels** | Oui | Oui | Moodle, Drupal, Joomla et connecteurs divers (contributions), Star, HAL, ESUP-ECM, Nuxeo |
| **Addons et extensions** | Une vingtaine (gratuite ou payante) | Environ 200 disponibles | Oui. La plupart des fonctionnalités courantes sont intégrées. |

D'une façon générale, la description sommaire et le tableau synthétique des trois outils conformes OAI-MPH montrent que DSpace et EPrints ont de nombreuses qualités, mais qu'ils ne sont pas adaptés au contexte français ou européen (respect des normes spécialisées, notamment pour les ressources pédagogiques et les thèses). En outre, ORI-OAI présente une grande souplesse de configuration et de personnalisation et s'adapte à tous les projets et à toutes les structures, du petit laboratoire jusqu'au regroupement d'universités et aux projets internationaux.

De par son origine française, plus adaptée au contexte des universités francophones, ORI-OAI se démarque ainsi comme alternative par défaut pour un portail commun de ressources éducatives francophones. Son application aux universités francophones serait plus économique en temps et en moyens puisque le processus d'extension ne demanderait qu'une mise à niveau et une adaptation des services documentaires dans les universités partenaires à un modèle partagé entre tous les acteurs concernés. L'expertise et le savoir-faire acquis par les UNT françaises (membres actifs du projet) dans la gestion des systèmes OAI-ORI, constitue un atout majeur à mettre au profit de la communauté éducative francophone.

La suite de cette étude sera donc concentrée sur les acquis d'un modèle de portail commun des ressources pédagogiques francophones conçu sur le modèle de fonctionnement d'ORI-OAI. En plus d'approfondir les avantages de cette solution, extraits de la littérature officielle du produit, nous passerons en revue son application sur les sites portails des UNT et les modalités de son optimisation pour un usage au sein d'un portail francophone commun.

## 6.2 ORI-OAI : CHOIX ADAPTE POUR UN PORTAIL FRANCOPHONE COMMUNICANT

En réalité, ORI-OAI n'est pas un outil de plus mais au contraire l'outil qui permet de faire le lien à la fois entre tous les autres outils et entre les différentes composantes d'un établissement ou d'un projet. Dans ce sens, ORI-OAI s'articule avec les outils présents dans les universités : il ne les remplace pas, mais il les valorise et démultiplie l'usage de leurs données par la mise en réseau.

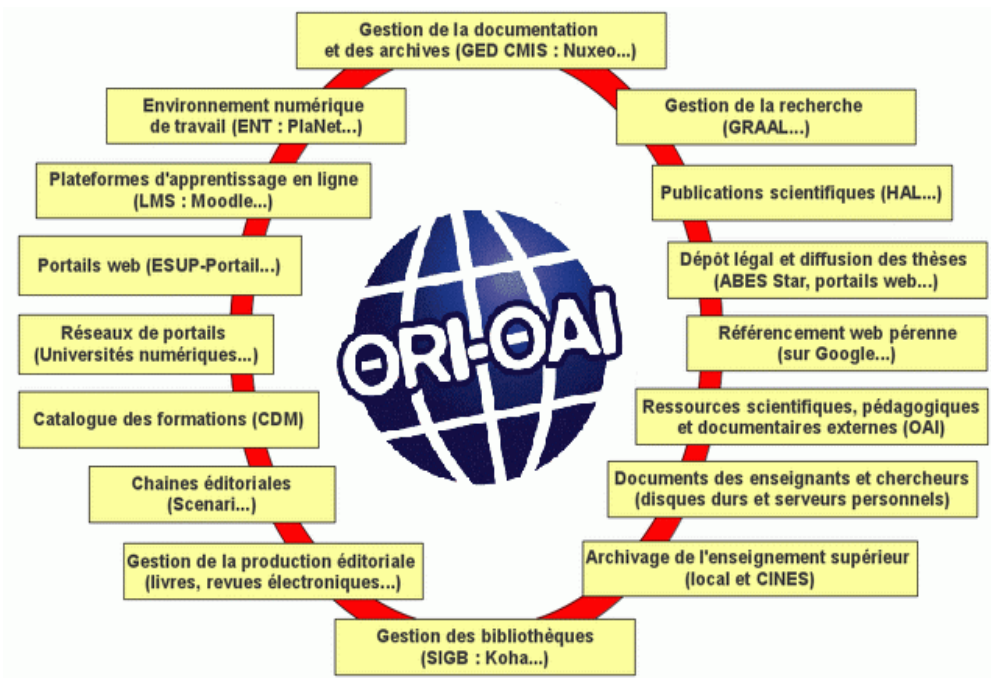

Source : *http://www.ori-oai.org/Positionnement*

ORI-OAI est donc le chaînon manquant entre les différentes composantes d'un établissement : il permet à plusieurs services ou même à plusieurs établissements de collaborer sur le référencement et l'indexation d'un document par le biais d'un circuit de traitement (workflow). D'autre part, ORI-OAI est le chaînon manquant qui permet la création et les échanges de métadonnées des ressources scientifiques, pédagogiques et documentaires gérées par les différents outils existants.

Sur le site de l'ORI-OAI (http://www.ori-oai.org/Positionnement), il est possible de mieux comprendre les formes d'articulation d'ORI-OAI avec les autres outils et services disponibles dans les établissements universitaires (Système Intégré de Gestion de Bibliothèque, gestion électronique des documents, archives ouvertes, SUDOC et le dépôt électronique des thèses, l'ENT, le LMS, la plate-forme pédagogique et la chaîne éditoriale, les outils de gestion de la recherche, les normes de référencement, l'indexation).

Une autre façon de comprendre ORI-OAI est de le considérer comme l'outil spécialisé dans la gestion cohérente et commune de tous les types de documents et de ressources des établissements de l'enseignement supérieur et de la recherche. Autrement dit, ORI-OAI est la fonctionnalité commune, mais sous-employée, de nombreux outils de l'enseignement supérieur : de fait, de nombreux outils gèrent des documents et disposent de fonctions basiques de référencement qui permettent de renseigner les métadonnées les plus courantes. Mais ces fonctionnalités sont souvent mal employées, car elles apparaissent secondaires à leurs utilisateurs compte tenu de la finalité même de ces autres outils, qui n'est pas le référencement. Surtout, les documents gérés par ces outils ne sont pas mis en valeur car ils ne sont connectés ni aux autres outils de l'établissement, ni aux réseaux externes, ce qui entraîne des coûts de gestion accrus et des pertes d'informations. ORI-OAI apporte donc de la cohérence dans le Système Global d'Information.

D'après ce constat (encadré), le modèle opératoire d'ORI-OAI ne présente aucune contrainte pour l'élaboration d'un portail communicant efficace. Il est suffisamment souple et varié pour permettre de gérer les différentes configurations des serveurs de données des universités francophones. C'est plutôt le sous-emploi des différentes fonctionnalités d'ORI-OAI qui expliquent les disparités et le développement inégal entre les services d'information des universités. Cette réalité est courante aussi bien en France que dans les pays francophones. Tout projet de fédération de services et ressources éducatives dans un portail francophone commun, devrait se concentrer d'abord sur les différentes formes d'articulation entre les services d'information existants dans les universités partenaires au moyen d'une mise à niveau ORI-OAI. Ce serait une phase préparatoire nécessaire avant l'intégration des serveurs institutionnels dans un portail francophone communicant de ressources pédagogiques numériques.

Cette phase passerait inéluctablement par l'observation des différents référentiels normatifs appliqués (et à recommander) pour l'harmonisation des cadres techniques et fonctionnels des différents partenaires du projet. S'agissant du partage de ressources numériques via un portail communicant, les référentiels normatifs de gestion, publication et valorisation des ressources pédagogiques devraient être a priori une préoccupation majeure pour les décideurs du futur portail francophone.

Dans ce sens, ORI-OAI permet de référencer les ressources pédagogiques des établissements en proposant l'utilisation de formats de métadonnées adaptés et normalisés (par des schémas de métadonnées et des profils d'application). Il permet également aux différents acteurs de l'établissement de participer à cette indexation (cellule TICE, SCD - Service Commun de la Documentation, service juridique...).

Dans le point suivant seront présentés les plus importants modèles de référencement normatif utilisés en faveur de la convergence des services et l'interopérabilité des ressources pédagogiques.

## 6.3 METADONNEES PEDAGOGIQUES POUR UN PORTAIL FRANCOPHONE

Comme tout autre domaine d'activité, l'éducation a su s'approprier les acquis technologiques pour se constituer un patrimoine normatif singulier. Plates-formes, vocabulaires, apprenants, supports de cours, offres de formation, accessibilité, etc. font désormais l'objet de normes internationales permettant des niveaux d'harmonisation nationales, régionales et internationales.

En France, les portails éducatifs se sont toujours alignés sur les référentiels de métadonnées innovants depuis l'émergence du format générique Dublin Core jusqu'au schéma plus récent du LOM (Learning Object Metadata) et ses profils d'application français « Lom.fr » et « SupLom.fr ». Aujourd'hui, ils sont encore appelés à s'aligner sur l'innovation qui sera apportée par le nouveau référentiel du MLR (*Metadata for Learning Resources*) en cours de publication par le Sous-comité 36 de l'ISO/IEC JTC1.

### 6.3.1 DUBLIN CORE

Dublin Core est un ensemble d'éléments simples mais efficaces pour décrire une grande variété de ressources en réseau. La norme du Dublin Core comprend 15 éléments dont la sémantique a été établie par un consensus international de professionnels provenant de diverses disciplines telles que la bibliothéconomie, l'informatique, le balisage de textes, la communauté muséologique et d'autres domaines connexes. Il s'agit des champs « Couverture », « Description », « Type », « Relation », « Source », « Sujet », « Titre », « Collaborateur », « Créateur », « Éditeur », « Droits », « Date », « Format », « Identifiant », « Langue ».

En tant que schéma de métadonnées, le Dublin Core est destiné à des ressources peu complexes et ne répond pas aux besoins de tous les métiers. On n'y trouve pas, par exemple des éléments de description propres au domaine de l'éducation. Pour cette raison on considère généralement qu'il appauvrit les données même si une ressource peut toujours être décrite en utilisant les propriétés du Dublin Core associé à d'autres propriétés venant d'autres vocabulaires ou ontologies.

Cependant, étant composé d'éléments génériques communs à presque tous les types de ressources, Dublin Core est repris au cœur de presque tous les schémas de métadonnées qui l'ont suivi. De ce fait, le Dublin Core reste bien un atout essentiel de l'interopérabilité et pas seulement dans le cadre de l'OAI-PMH.

### 6.3.2 LOM

Le LOM est un standard international proposant un modèle de description des métadonnées associées à des objets pédagogiques quels qu'ils soient, numériques ou non. Il ne constitue pas une norme en soi, mais plutôt une recommandation de métadonnées. Il a été adopté par les IEEE en 2002 pour proposer un jeu de métadonnées découpé en neuf catégories : général, cycle de vie, méta métadonnées, technique, pédagogique, relations, droits, commentaires, classification.

La catégorie 5 (pédagogie) constitue la différence principale qui inscrit le LOM dans le domaine de l'éducation. Depuis lors, le LOM devient le schéma de métadonnée de référence pour les profils d'application nationaux qui se sont appropriés ce modèle normatif et l'ont adapté à leurs besoins locaux.

En France, l'adaptation du LOM a donné lieu à deux profils d'application, l'un national (LOM.FR) et l'autre encore plus spécifique au domaine universitaire (SupLOM.FR).

Sur les portails éducatifs en France, lors de l'installation d'ORI-OAI, trois formats de métadonnées sont proposés pour la description des ressources pédagogiques : le LOM et ses deux déclinaisons françaises, LOMFR et SupLOMFR.

### 6.3.3 LOMFR

Contrairement au LOM, le profil français d'application du LOM (LOMFR) est une norme élaborée par le Ministère de l'Education nationale et l'AFNOR en 2006 pour décrire des ressources pédagogiques produites par l'ensemble de la communauté éducative française.

Selon EducNet, « Le LOMFR décrit des objets (ressources) pédagogiques. Est considérée comme ressource pédagogique toute entité (numérique ou non) utilisée dans un processus d'enseignement, de formation ou d'apprentissage. Il peut s'agir de livres, de films, de scénarios pédagogiques, de sites web, de logiciels, etc. Ses objectifs sont les suivants : permettre une meilleure visibilité de l'offre d'objets pédagogiques, numériques ou non ; disposer de normes à respecter pour garantir la pérennité, l'interopérabilité, et une certaine ouverture des ressources numériques produites, en lien avec les plates-formes de distribution et de consultation de ces ressources (Extrait de la norme expérimentale). Il concerne l'ensemble de la communauté éducative : communauté éducative française (enseignements primaire, secondaire, supérieur) ; communauté de la formation (formation professionnelle, formation continue) ; documentalistes et bibliothécaires ; éditeurs, entreprises. » (Source : site EducNet - http://www.educnet.education.fr/dossier/metadata/lom1.htm)

L'objectif de LOMFR est de fournir un cadre de travail à la fois cohérent avec le système éducatif français et conforme avec les règles de description définies au niveau international.

### 6.3.4 SUPLOMFR

SupLOMFR est une adaptation de LOMFR lui-même décliné de LOM. Les ministères de l'Education nationale et de l'Enseignement supérieur et de la recherche ont souhaité adapter la norme LOMFR pour tenir compte des besoins et contraintes spécifiques des établissements d'enseignement supérieur et des universités numériques tout en restant conforme aux préconisations nationales.

Un groupe de travail SupLOMFR, composé des représentants du ministère et des établissements, s'est réuni à partir de 2007 pour définir d'une part le caractère obligatoire ou recommandé des éléments du LOMFR et d'autre part les nomenclatures, classifications et vocabulaires préconisés.

SupLOMFR est une adaptation de LOMFR pour tenir compte de besoins et contraintes spécifiques aux établissements d'enseignement supérieur, aux universités numériques thématiques, et aux universités numériques en région tout en restant conforme aux préconisations nationales. Ce travail a été souhaité par les ministères de l'Education nationale et de l'Enseignement supérieur et de la recherche, dans le cadre de leur soutien à l'édition numérique pour l'éducation.

## 6.3.5 MLR

Le MLR (*Metadata for Learning Resources*) est une nouvelle norme pédagogique initiée et produite au sein du SC36 de l'ISO. Publiée initialement en 2011, elle a été soutenue par la délégation française de l'Afnor pour faire face à une perte progressive de l'interopérabilité pédagogique internationale issue d'une prolifération de profils d'application LOM et d'autres schémas totalement indépendants. Selon Bourda et Delestre[5], « l'interopérabilité entre ces schémas n'est plus automatique. Pour résoudre ce problème, nous proposons d'utiliser une méthode définie par la norme ISO11179 qui permet de dissocier l'aspect conceptuel de l'aspect représentation lors de la conception de schéma de métadonnées. Ainsi, grâce à ce modèle, les schémas de description de documents pédagogiques pourraient être considérés comme des instanciations d'un seul et unique modèle conceptuel, et donc devenir interopérables ».

« Pour que les ressources pédagogiques décrites par les différents profils LOM soient réutilisables, il faut que les métadonnées les décrivant soient organisées selon un schéma interopérable avec tous les autres. L'interopérabilité doit s'entendre au sens qu'une entité pédagogique, référencée dans un système, peut être réutilisée par un autre système possédant un schéma de métadonnées différent du premier. Le format MLR (metadata for learning resources) mis au point par l'ISO SC36, est un modèle de conception de schéma de métadonnées pédagogiques conceptuelles. Son rôle est de rendre un schéma de métadonnées interopérable avec un autre, grâce à son niveau conceptuel et donc générique. Pour cette raison, le MLR peut être la solution au problème d'interopérabilité entre ressources pédagogiques »[6].

Actuellement, beaucoup d'initiatives sont déjà en cours de réalisation pour préparer la migration du LOM (et ses profils d'application) vers la nouvelle norme MLR dans divers profils d'application. A titre d'exemple, le GTN-Québec développe une nouvelle version de son profil NORMETIC fondée sur cette optique. Alors que l'actuelle version est un profil d'application du IEEE LOM la nouvelle version, Normetic 2.0, sera basée sur la norme internationale ISO/IEC 19788-1:2011. L'objectif du projet est de fournir une première version d'un outil permettant de transformer une fiche LOM (format XML – IEEE Std 1484.12.3-2005) en un enregistrement MLR (format RDF, syntaxe Turtle)

Au niveau français, une pareille initiative est encore à développer au sein des portails éducatifs pour optimiser les mécanismes de référencement et de conversion des ressources pédagogiques depuis les profils d'application LOMFR et SUPLOMFR actuellement d'usage vers un profil actualisé inspiré des directives du MLR.

---

[5] Bourda Y, Delestre N. « Améliorer l'interopérabilité des profils d'application du LOM ». Revue STICEF. Vol. 12, 2005.

[6] Bentaieb Amine et Arnaud Michel. Le projet de format ISO SC36 MLR (metadata for learning resources) peut rendre interopérables les profils d'application LOM. www-sop.inria.fr/acacia/...List/WebLearn05-Bentaieb-Arnaud-1.doc

### 6.3.6 UN FORMAT FRANCOPHONE UNIFIE DE METADONNEES PEDAGOGIQUES

Un portail communicant de ressources pédagogiques francophones gagnerait beaucoup à s'inscrire d'emblée dans cette prospective normative, surtout que l'opération ne consiste pas à réinventer un nouveau format mais plutôt à développer une application de migration automatique d'un profil LOMFR à un profil francophone MLR.

Une étude des caractéristiques d'un format commun de métadonnées francophone devrait être lancée d'urgence pour proposer un cadre alternatif de référencement des ressources pédagogiques qui serait proposé aux partenaires universitaires francophones (cf. recommandations).

La question à poser pour les universités francophones partenaires serait de savoir s'il faudrait d'abord développer des profils d'applications conformes à leurs spécificités locales en adaptant un LOM national (voire sectoriel) ou de s'aventurer d'emblée dans un projet de format francophone commun de métadonnées pédagogiques sur la base d'un schéma MLR qui monte en puissance. Une étude préalable devrait être élaborée pour déterminer les effets escomptés et les acquis engendrés par une démarche ou une autre.

A la lumière des éléments de description dans les points précédents qui ont permis d'exposer le contexte général de l'étude et le contexte technologique présidant au développement et usage des portails des ressources pédagogiques numériques, il serait opportun de procéder à une description de l'état de l'art des portails des institutions concernées par la commande DNEUF. Pour ce faire, on définira un ensemble de critères qu'on confrontera de façon sommaire à un ensemble de portails référents, notamment les portails des UNT et quelques portails référents internationaux.

# 7 EVALUATION DES PORTAILS : CRITERES ET MODELE DE DESCRIPTION

Pour faire l'étude d'un modèle de portail de ressources pédagogiques utiles à la commande DNEUF, il est nécessaire de partir d'un ensemble de critères de description (et d'évaluation) qui focaliserait les caractéristiques jugées utiles pour la construction du portail des ressources numériques dans l'espace universitaire francophone. Ce jeu de critères concernerait normalement autant les techniques d'information (classification, description, indexation des ressources,…) que les modes de gestion (organisation, gestion, suivi, mise à jour, …) et de communication (interfaces, accessibilité, échange, diffusion,…). Mais dans cette étude, nous limiterons notre intérêt aux critères à portées documentaire et pédagogique, proposant de reporter l'analyse ergonomique des interfaces de navigation (chartes graphiques et navigationnelle) à des études ultérieures qui prépareraient à la réalisation concrète du portail.

Les critères retenus seront appliqués à un ensemble de portails significatifs par leurs fiabilités fonctionnelles et leurs notoriétés institutionnelles. Il existe en effet de nombreux « faux » portails éducatifs sur le web. Certains pourraient être classés comme dépôts de ressources sans aucun service à valeur ajoutée (diffusion de contenus) ; d'autres sont de simples espaces sur le web où les données provenant d'autres sites peuvent être stockées. Habituellement, ces sites ne fournissent pas des informations à jour et la plupart du temps ils sont sans aucune sorte d'interaction avec les utilisateurs.

Pour le besoin de cette étude, nous avons optés pour les sites des universités Numériques Thématiques en France et quelques sites internationaux référents pour leurs fiabilités et réputations confirmées. Nous avons d'abord procédé à une identification d'un ensemble d'indicateurs qui, de notre point de vue, renvoient aux valeurs génériques de qualité qui doivent nécessairement figurer dans un portail de ressources numériques. Ces critères concernent :

- *La facilité d'utilisation* : les portails doivent être faciles de compréhension. L'utilisateur doit être autonome et les icônes et symboles doivent correspondre aux services offerts afin de faciliter sa manipulation.

- *Les services d'assistance* : le portail doit offrir des services d'aide comme les FAQ (Foire Aux Questions), le téléchargement de logiciels nécessaires à l'exécuter ou la visualiser du contenu (plugins, lecteur PDF, visionneuses vidéo, etc.).

- *Les services de communication* : le portail doit offrir des outils de communication synchrones (chat) et des outils de communication asynchrone (forum, e-mail, lettre d'information, liste de diffusion), pour la communication entre les utilisateurs eux même et avec l'administrateur du portail.

- *Le contenu* : comme il s'agit d'un portail éducatif, il devrait nécessairement offrir à ses utilisateurs des contenus et des ressources didactiques dans des formats variés. Pour cela, il devrait disposer de ressources multimédia (texte, image, son, vidéo et animation). Ce critère est sans aucun doute celui qui doit recevoir la plus grande attention de la part des administrateurs de portail, car, même avec peu de services et avec peu de convivialité, il est possible qu'un site réussisse sa mission pédagogique de transmission de connaissances comme le font à certaines limites certains blogs.

- *La performance* : le portail devrait offrir un temps satisfaisant de chargement de ressources. L'utilisateur ne devrait pas avoir un délai d'attente assez long pour visualiser le contenu. Ce critère peut compromettre le succès du portail.

- *L'information* : en tant que service minimum, un portail éducatif devrait offrir une information mise à jour, si possible quotidiennement, à partir de sources fiables. C'est l'un des facteurs qui peuvent faire revenir les utilisateurs du portail.

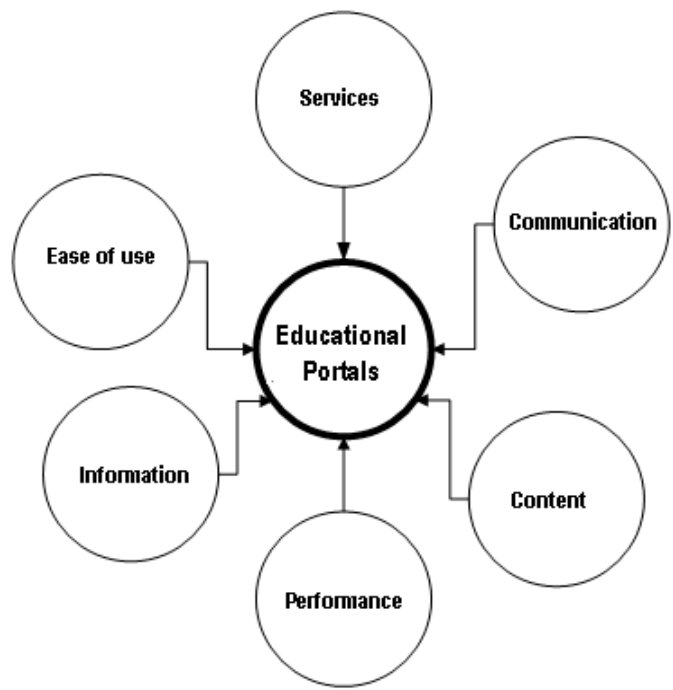

Critères génériques de qualité d'un portail éducatif

Pour nous concentrer sur une analyse portée sur les spécificités pédagogiques des portails référents, il a fallu décliner ces critères génériques en une série de questions ciblées que nous regroupons sous cinq facettes : catégorisation (cadre général du portail), organisation de ses ressources pédagogiques, leur description, leur traitement et leur exploitation.

❖ **Catégorisation (cadre général du portail)**

L'identification des portails des ressources pédagogiques est relevée dans cette étude à travers les critères suivants :

- Nom/sigle du portail ;
- Nature du portail (préciser s'il s'agit d'un portail du secteur public, privé ou associatif) ;
- Mode d'accès aux ressources («gratuit» ou «payant» ; «ouvert» ou «sur inscription») ;
- Tutelle et/ou porteur du portail ;
- Population cible (distinguer si le portail est pour étudiants et/ou enseignants et/ou autres) ;
- Aspects linguistiques (ressources multilingues) ;
- Nombre de ressources pédagogiques accessibles ;
- Niveau académique des ressources disponibles (collège, cycles LMD, formation professionnelle, etc.) ;
- Typologie informatique des ressources disponibles (fichiers texte, multimédias)
- Typologies éducative des ressources pédagogiques (cours, exercices, démos, etc.) ;

❖ **Organisation**

L'organisation des ressources pédagogiques sur un portail institutionnel peut prendre plusieurs formes. Elle est identifiée ici à travers les critères suivants :

- Les ressources sont disposées en collections (thématiques, niveaux scolaires, types de supports, etc.) ;
- Le portail dispose de sa propre base de ressources (réservoir d'objets pédagogiques) accessibles par ses propres outils de recherche (annuaire et moteur) [vérification par URL] ;
- Le portail dispose d'une base de référence (de métadonnées) liée à des ressources hébergées sur des serveurs externes ;
- Le portail ne dispose ni bases de références ni de base de ressources ni d'outils de recherche propres mais propose une liste de signets vers des ressources pédagogiques sur des sites externes ;
- L'alimentation du portail en ressources pédagogiques est soit réservée à un personnel spécialisé (auteurs et documentalistes) ou ouverte au public (sur validation préalable) ;
- Données sur les droits d'auteurs des ressources pédagogiques.

❖ **Description/moissonnage**

La description des ressources pédagogiques est un processus documentaire qui consiste à décrire les documents par des mots clés (métadonnées) utiles autant pour l'identification que l'indexation et la recherche. La description s'effectue généralement par des éléments de métadonnées fournis par l'auteur (saisies au moment de la création de la ressource), produites par le système d'information (date, format, taille), ou ajoutées a posteriori par des documentalistes pour des finalités d'optimisation de la recherche. Ces métadonnées sont moissonnées par les moteurs de recherche pour créer les bases de références (index) au profit de la recherche. Parmi les éléments à vérifier sous la facette de la description :

- Résumés (vérifier si les ressources sont accompagnées de résumés synoptiques) ;
- Métadonnées (vérifier si les ressources sont décrites par des métadonnées) ;
- Critères de description pédagogique utilisés (niveau scolaire, durée d'apprentissage, méthode d'apprentissage, type d'évaluation, etc.) ;
- Schéma de métadonnées ou profil d'application utilisé (DC, LOM, LOMFR, SupLOMFR, MLR ou format libre) ;
- Vocabulaire ou langage contrôlé (index, listes de choix, taxonomie, thésaurus, etc.) ;
- Moissonnage OAI-PMH ou autre ;

❖ **Exploitation**

L'exploitation des ressources pédagogiques fournit autant à l'utilisateur final qu'à l'administrateur du portail un certain nombre de fonctionnalités capables de faciliter l'accès aux collections de documents et leur exploitation pour créer de la valeur ajoutée. Parmi les services à observer :

- Double accès aux ressources en mode annuaire et en mode recherche locale ;
- Ressources accessibles par les moteurs de recherche génériques (Test Google) ;
- Recherche par filtrage multicritère (cohage par catégories) ;
- Recherche booléenne (et, ou, sauf) ;
- Services de veille (alertes, RSS…) ;
- Outils de partage Web 2.0 (Facebook, Twitter, etc.)
- Affichage automatique de listes de nouveaux contenus ;
- Reconnaissance automatique des ressources par les outils bibliographiques (Test Zotero) ;
- Exportation des références bibliographiques (formats Zotero, CSV, XML, etc.) ;
- Connecteurs à des applications externes (Moodle, etc.) ;

Comme nous l'avons déjà signalé, ces critères sélectifs sont d'ordre à fournir une idée sur la structuration et l'usage des ressources pédagogiques et les techniques documentaires qui leur sont appliquées sur les portails étudiés. Une évaluation ergonomique étendue de ces portails est toujours possible par d'autres critères comme la disponibilité d'une FAQ (*Frequently Asked Questions*), d'enquêtes de satisfaction en ligne, de formulaires de suggestions, de compteurs de visites géo-localisées, de menu de navigation toujours visible, de plan de portail, de chemin de navigation (fil d'Ariane), d'enregistrement d'utilisateurs (notion de communauté), d'outils de communication (messagerie, réseaux sociaux), etc.

# 8 ANALYSE DE PORTAILS DE RESSOURCES PEDAGOGIQUES

Nous entamerons dans cette section du rapport la description de portails éducatifs francophones jugés représentatifs d'une politique éducative du numérique avant-gardiste en France et de quelques portails de ressources éducatives de grosses universités dans le monde. La finalité première en est surtout de repérer des tendances reproductibles dans la conception et gestion de portails pédagogiques de qualité. Les constats seraient utiles pour déterminer la suite de la stratégie d'action dans la construction du portail commun des ressources pédagogiques francophones.

Répondant à un critère de la commande DNEUF exposé en introduction, celui d'étendre aux institutions universitaires francophones les capacités du moteur de recherche de France Université Numérique, notre analyse portera essentiellement sur les portails des UNT comme modèles reproductibles ou extensibles au contexte francophone. Nous élargissons ensuite l'étude d'autres portails existant au niveau mondial pour explorer d'autres modèles d'organisation de portails de ressources pédagogiques.

## Portails francophones

En Francophonie, plusieurs expériences de portails de ressources pédagogiques sont recensées. Elles se concentrent essentiellement entre la France et le Canada.

Parmi les modèles français en matière de portails communicants de ressources pédagogiques, ceux du portail du numérique dans l'enseignement supérieur (sup-numerique.gouv.fr), du portail France Université Numérique - Mooc (FUN-MOOC) et des Universités Numériques Thématiques (UNT) sont les plus indiqués.

Sur le plan international, le Canada, tout en proposant des expériences de portails francophones, développe aussi dans ses espaces académiques anglo-saxonnes des expériences de portails intéressants à étudier. En Europe ou aux Etats-Unis d'Amérique, les expériences ne sont pas moindres.

## 8.1 LE PORTAIL DU NUMERIQUE DANS L'ENSEIGNEMENT SUPERIEUR (SUP-NUMERIQUE.GOUV.FR)

Le portail du numérique dans l'enseignement supérieur (sup-numerique.gouv.fr) s'inscrit dans le cadre de l'extension de la stratégie numérique pour l'enseignement supérieur, mise en œuvre depuis 2013 par le Ministère de l'Education nationale, de l'Enseignement supérieur et de la Recherche. Ce nouveau portail constitue un guichet unique dédié à l'enseignement supérieur par le numérique. Il se substitue au portail France Université Numérique (FUN) qui a fait l'objet d'une large refonte pour devenir le portail sup-numerique.gouv.fr[7].

Le portail sup-numerique.gouv.fr s'articule surtout avec les Universités Numériques Thématiques (UNT) et Canal-U ainsi que la plate-forme FUN-MOOC. Pour entretenir cette nouvelle organisation et la conduire dans un cadre de cohérence et de convergence avec la politique éducative des institutions, le portail du numérique dans l'enseignement supérieur (sup-numerique.gouv.fr) est mis en synergie avec les services documentaires des établissements de l'enseignement supérieur français et de leurs partenaires académiques dans le monde entier.

Outre une arborescence globale entièrement revue, le portail sup-numerique.gouv.fr propose une meilleure hiérarchisation de l'information : les éléments pratiques sont désormais facilement identifiables, la navigation au sein des pages est facilitée (navigation d'un contenu à l'autre dans la même rubrique et sur les mêmes thèmes). Le portail est conçu pour les utilisateurs qui peuvent donner leur avis et participer activement à l'amélioration du site spécialement sur les aspects des contenus, les aspects techniques et les fonctionnalités et services proposés. En étant adaptatif, il est consultable sur ordinateurs, téléphones et tablettes. Les possibilités de partage sont renforcées, des Twitcards et posts Facebook sont embarqués à chaque partage[8].

Au cœur du portail du numérique dans l'enseignement supérieur (sup-numerique.gouv.fr) deux types de moteurs de recherche facilitent l'accès à deux catégories de données : le contenu du site hors ressources pédagogiques (pages d'information) et le contenu des ressources pédagogiques.

Cette répartition est provisoire dans l'attente d'une solution unique optimisée pour la recherche de tous types de ressources.

✓ *Le moteur dédié aux contenus du site, hors ressources pédagogiques*

Ce moteur est basique. Il est fondé sur mode de recherche simple sans alternance avec un mode de recherche avancée. En revanche, il met en application un ensemble de filtres orientés vers les types de contenus recherchés, les thèmes de recherche, des intervalles de dates (sans préciser de quel type), un ordre de pertinence limité à un seul critère de date et enfin un filtre de nombre de résultats affichées par page.

---

[7] Il est à rappeler à titre de précision que cette évolution était liée au fait que le portail et la plateforme de MOOC (France Université Numérique) portaient le même nom et la même identité visuelle. L'arbitrage ministériel au printemps 2015 a visé à supprimer la confusion engendrée pour les internautes en rebaptisant le portail (qui n'utilise plus FUN) et en laissant l'usage de FUN à la seule plateforme de MOOC (FUN-MOOC) devenue un sous service du portail sup-numerique.gouv.fr.

[8] Allocution de Thierry Mandon « Lancement du portail sup-numerique.gouv.fr », Communiqué - 22.10.2015 sur http://www.enseignementsup-recherche.gouv.fr/cid94588/lancement-du-portail-sup-numerique.gouv.fr.html

**Type de contenu**

Actualité
Appel à projets et à candidatures
Article
Communiqué
Dossier de presse
MOOCs
Publication
Événements, manifestations

**Filtre thématique**

Thèmatique
4 ambitions
    Développement de campus d'avenir
    Europe et international
    Rénovation des pratiques pédagogiques
    Réussite et insertion des étudiants
Autres organismes
    Conférence des présidents d'université (C.P.U.)
Autres établissements
    Alliance française Paris Ile-de-France
    Animafac

✓ *Le moteur dédié aux ressources pédagogiques*

La simplicité du moteur de recherche hors ressources pédagogiques est compensée par des fonctionnalités plus avancées au niveau du moteur des ressources pédagogiques. La fusion prévue des deux moteurs convergera inéluctablement vers une solution optimisée avec les possibilités de recherche proposées par le moteur des ressources pédagogiques. Ce dernier est basé sur un mode de recherche par mots clés exclusifs (tous les mots), inclusifs (un des mots) ou par « expression exacte ». Il met aussi en application un ensemble de filtres orientés vers les disciplines, les types de ressources, le niveau de scolarité, le format physique et l'âge de la ressource.

- Recherche par discipline

- Recherche par type de ressource

- Recherche par niveau de scolarité

| RECHERCHER PAR | | | | |
|---|---|---|---|---|
| Discipline ⌄ | Type de ressource ⌄ | Niveau ⌄ | Format ⌄ | Age de la ressource ⌄ |
| ☐ enseignement supérieur (**33095**) ✚ | | ☐ formation professionnelle (**5496**) ✚ | | |

- Recherche par Format

| RECHERCHER PAR | | | | |
|---|---|---|---|---|
| Discipline ⌄ | Type de ressource ⌄ | Niveau ⌄ | Format ⌄ | Age de la ressource ⌄ |
| ☐ vidéo (**19892**) | | ☐ texte (**12497**) | | ☐ ressource interactive (**2673**) |
| ☐ image (**1923**) | | ☐ image fixe (**1350**) | | ☐ son (**1042**) |
| ☐ ensemble de données (**226**) | | ☐ collection (**169**) | | ☐ logiciel (**40**) |
| ☐ objet physique (**7**) | | ☐ évènement (**3**) | | |

- Recherche par Age de la ressource

| RECHERCHER PAR | | | | |
|---|---|---|---|---|
| Discipline ⌄ | Type de ressource ⌄ | Niveau ⌄ | Format ⌄ | Age de la ressource ⌄ |
| ☐ moins d'un an (**670**) | | ☐ de 1 à 2 ans (**2633**) | | ☐ de 2 à 5 ans (**9120**) |
| ☐ plus de 5 ans (**17681**) | | | | |

Le moteur des ressources pédagogiques a été lancé en juin 2015 pour donner accès aux ressources mises à disposition par les établissements partenaires, notamment les UNT et Canal-U. Ce moteur de recherche donne un accès gratuit et rapide aux ressources pédagogiques avec une sélection plus précise. Il est ouvert aux étudiants, aux enseignants, aux chercheurs, aux professionnels et plus généralement au grand public pour accéder à plus de 30.000 ressources pédagogiques numériques sous forme de cours, études de cas, tutoriels, leçons interactives, conférences, proposées sous forme textes, vidéos, webdocumentaires, logiciels ou sites internet.

Sur le plan technique, le moteur du portail sup-numerique.gouv.fr a été développé pour répondre à quatre exigences principales :

1. Référencer les ressources existantes et créer un point d'accès unique ;

2. Faciliter l'accès de tous aux ressources pédagogiques en ligne, simplement et gratuitement grâce à des outils de recherche performants ;

3. Faire remonter des résultats de recherche pertinents ;

4. Permettre aux établissements d'enseignement supérieur, aux organismes de recherche et aux U.N.T. de valoriser et de partager leurs travaux.

Pour être conforme à ces exigences, le moteur du portail sup-numerique.gouv.fr a été conçu selon les principes opératoires du projet ORI-OAI (Outil de Référencement et d'Indexation – Réseau de Portail OAI) et les règles du protocole OAI-PMH (*Open Archive Initiative - Protocol for Metadata Harvesting* =

protocole de collecte de métadonnées pour l'initiative des archives ouvertes). Ce protocole mondialement connu dans l'échange et le transfert, est basé sur les métadonnées sans avoir besoin de rassembler physiquement les ressources indexées. En effet, le protocole OAI-PMH facilite la mutualisation et la diffusion des contenus car il repose sur l'utilisation des fiches de métadonnées.

A ce titre, les ressources intégrées au portail, pour qu'elles soient accessibles et interopérables, ont été décrites grâce au schéma SupLOMFR, un schéma de métadonnées dérivé du LOM et du LOMFR et spécifiquement adapté à l'enseignement supérieur. L'indexation du moteur a toujours été disponible sous format de jeu de données Open Data.

Dans la mouvance internationale du monde éducatif qui progresse lentement vers MLR (Metadta for Learning Resources), un substitut au schéma générique LOM (Learning Object Metdata) et ses profils d'application nationaux, le projet ORI-OAI (et par conséquent le portail sup-numerique.gouv.fr) est supposé accomplir dans les brefs délais une migration vers un nouveau modèle de métadonnées pédagogiques, en l'occurrence un profil d'application français du schéma MLR (*Metadata for Learning Resource*s). Comme mentionné précédemment, le M.L.R. est une nouvelle norme innovante en plusieurs parties dont l'objectif est la description des ressources pédagogiques dans un contexte international multilingue et multiculturel. Cette norme prend en compte les dernières évolutions en termes de diffusion de données ouvertes et liées et reste indépendante de toute technologie.

✓ **Catégorisation (données d'identité)**

| | |
|---|---|
| Nom/sigle du portail ; | 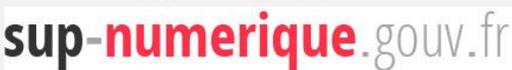<br>http://www.sup-numerique.gouv.fr/ |
| Nature du portail (préciser s'il s'agit d'un portail du secteur public, privé ou associatif) ; | Secteur public (MESRS) |
| Mode d'accès aux ressources («gratuit» ou «payant» ; «ouvert» ou «sur inscription») ; | Ouvert et gratuit (avec deux espaces différents, l'un aux ressources pédagogiques et l'autre aux MOOCs) |
| Tutelle et/ou porteur du portail | MESR |
| Population cible (distinguer si le portail est pour étudiants et/ou enseignants) ; | Etudiants, enseignants, chercheurs, professionnels et plus généralement grand public |
| Aspects linguistiques (contenus multilingues) ; | Multilingue (anglais/français) |
| Nombre de ressources pédagogiques accessibles ; | 33807 (à la date du 03/02/2016) |

| Niveau académique des ressources disponibles (collège, cycles LMD, formation professionnelle, etc.) ; | Enseignement supérieur, formation professionnelle |
|---|---|
| Typologie informatique des ressources disponibles (fichiers texte, multimédias) | textes, vidéos, logiciels, sites internet |
| Typologies éducative des ressources pédagogiques (cours, exercices, démos, etc.) ; | cours, études de cas, tutoriels, leçons interactives, conférences, proposées sous forme textes, vidéos, web documentaires, logiciels ou sites internet |

- ✓ **Organisation**

| Les ressources sont disposées en collections (thématiques, niveaux scolaires, types de supports, etc.) ; | disciplines, types de ressources, niveau de scolarité, format physique et âge de la ressource |
|---|---|
| Le portail dispose de sa propre base de ressources (réservoir d'objets pédagogiques) accessibles par ses propres outils de recherche (annuaire et moteur) [vérification par URL] ; | Non/oui :<br>- Des ressources pédagogiques propres aux sites partenaires accessibles par annuaire et moteur de recherche,<br>- Des cours en lignes (Moocs) hébergés sur des serveurs MOOCs externes |
| Le portail dispose d'une base de référence (de métadonnées) liée à des ressources hébergées sur des serveurs externes ; | Oui (fiches de métadonnées) |
| Le portail ne dispose ni de bases de références ni de base de ressources ni d'outils de recherche propres mais propose une liste de signets vers des ressources pédagogiques sur des sites externes ; | Non : portail avec un système d'information propre et des services à valeurs ajoutées |
| L'alimentation du portail en ressources pédagogiques est soit réservée à un personnel spécialisé (auteurs et documentalistes) ou ouverte au public (sur validation préalable) ; | - Partie ressources : Gérée au niveau des institutions partenaires<br>- Partie Moocs : cours proposés par les institutions avec l'appui de référentes (1 par site) et correspondants (1 par établissement) |
| Données sur les droits d'auteurs des | Non (héritées du serveur partenaire et de l'auteur de la |

| ressources pédagogiques. | ressource) |
|---|---|

- ✓ **Description/moissonnage**

| Résumés (vérifier si les ressources sont accompagnées de résumés synoptiques) | Oui |
|---|---|
| Métadonnées (vérifier si les ressources sont décrites par des métadonnées) | Oui |
| Critères de description pédagogique utilisés (niveau scolaire, durée d'apprentissage, méthode d'apprentissage, type d'évaluation, etc.) | Niveau, type de contenu, type de document, poids, durée d'exécution |
| Schéma de métadonnées ou profil d'application (DC, LOM, LOMFR, SupLOMFR, MLR ou format libre) | LOMv1.0, LOMFRv1.0 (Modèle de données basé sur schéma LOM) |
| Vocabulaire ou langage contrôlé (index, listes de choix, taxonomie, thésaurus, etc.) ; | Index alphabétique (interactif) sur zone recherche |
| Moissonnage OAI-PMH ou autre ; | OAI-PMH (ORI-OAI) |

- ✓ **Exploitation**

| Mode d'accès aux ressources («gratuit» ou «payant» ; «ouvert» ou «sur inscription») | Ouvertes et gratuites (Moocs : gratuits sur inscription) |
|---|---|
| Double accès aux ressources en mode annuaire et en mode recherche locale ; | Oui |
| Ressources accessibles par les moteurs de recherche génériques (test Google) ; | Oui |
| Recherche par filtrage multicritère | Oui (Discipline, type, niveau, format) |
| (mode cochage par catégories) ; | |
| Recherche booléenne (et, ou, sauf) | Oui (implicite : « tous les mots », « un seul mot », « expression exacte ») |
| Services de veille (alertes, RSS…) | Non |
| Outils de partage Web 2.0 (Facebook, Twitter, etc.) | Twitter, Facebook, Viadeo, LinkedIn, messagerie |
| Affichage automatique de listes de nouveaux contenus ; | Nouvelles ressources |
| Reconnaissance automatique par les outils bibliographiques (Test Zotero) | Non |
| Exportation des références bibliographiques (formats Zotero, CSV, XML, etc.) | XML (fiches de métadonnées) |
| Connecteurs à des applications externes (Moodle, etc.) | Non |

On voit bien à travers ce bref relevé de caractéristiques techno-documentaires que le portail sup-numerique.gouv.fr accumule une double fonction de fournisseur de données et fournisseurs de services.

Rappelons que les fournisseurs de données (data providers) sont responsables des réservoirs d'information (documents primaires : archives ouvertes, dépôts) qui centralisent les données et peuvent éventuellement donner accès au texte intégral. Ils appliquent le protocole de collecte de métadonnées des archives ouvertes OAI-PMH afin de mettre en valeur les métadonnées du contenu de leurs réservoirs de données.

Rappelons aussi que les fournisseurs de services (service providers) récoltent les métadonnées à partir des interfaces OAI (ou portails) des fournisseurs de données pour collecter et centraliser les métadonnées. Ils offrent des services à valeur ajoutée. Ils peuvent enrichir eux-mêmes les métadonnées ainsi récoltées.

Selon cette distinction, sup-numerique.gouv.fr fait fonction de fournisseur de données grace à son propre réservoir de cours en version Mooc qu'il soumet aux opération d'indexation par ORI-OAI en open data. Il fait aussi fonction de fournisseur de services en collectant les métadonnées des portails des institutions partenaires (nortamment les UNT et Calal-U) tout en les amélirant pour en produire des services à valeurs ajoutées accessibles sur son propre serveur.

# FRANCE UNIVERSITE NUMERIQUE (FUN-MOOC)

Devenue un service intégré au portail sup-numerique.gouv.fr, la plate-forme FUN-MOOC (France Université Numérique - Massive Open Online Courses) rejoint d'autres plates-formes (auto-hébergé, sur Canvas, CNC, Eco-Learning, IONISx, Open Clasroom, Unow, etc.) pour donner accès à un grand nombre de formations MOOC. L'accès au contenu de ces plates-formes ne s'effectue pas encore par un moteur de recherche propre au portail sup-numerique.gouv.fr mais se limite aux services du moteur de recherche proposé par chaque plate-forme visitée.

Le portail sup-numerique.gouv.fr propose en revanche le filtrage des Moocs recensés par les noms des plates-formes qui les hébergent et par discipline.

Dans cet agrégat de plates-formes MOOC, FUN-MOOC est présentée comme l'opérateur du portail FUN. Il s'agit d'un Groupement d'Intérêt Public (GIP) cofinancé par ses établissements membres (CINES, Inria et Renater) et le Ministère de l'Enseignement Supérieur et de la Recherche. Il s'agit d'une plate-forme présentée via un portail web ouvert et gratuit. Ce portail a pour vocation de permettre aux universités et écoles françaises qui le souhaitent de dispenser leurs cours en ligne via la plate-forme FUN-MOOC. Initialement lancée par le Ministère de l'Enseignement Supérieur et de la Recherche en octobre 2013, cette initiative vise à fédérer les projets des universités et écoles françaises pour leur donner une visibilité internationale, et permettre à tous les publics d'accéder à des cours variés et de qualité où qu'ils soient dans le monde. FUN-MOOC dispose donc d'un réservoir important de ressources pédagogiques qui, proprement dit, sont hébergées par une plate-forme distincte du portail et accessible gratuitement par tous après inscription.

## 8.2 UNIVERSITES NUMERIQUES THEMATIQUES (UNT)

Les UNT (Universités Numériques Thématiques) s'inscrivent aussi dans la politique française de développement des ressources pédagogiques numériques. Elles ont pour mission, dans le cadre d'une mutualisation à une échelle nationale, de recenser les ressources pédagogiques numériques existantes dans les établissements, de produire de nouvelles ressources, de valoriser, indexer et diffuser ces ressources et d'assurer la validation scientifique, pédagogique et technique des ressources produites.

Les UNT mutualisent, à l'échelle nationale, des contenus pédagogiques produits par des enseignants des établissements d'enseignement supérieur français, de toute nature (documents, cours, exercices, exemples, etc.), dans tout domaine disciplinaire et pour toute forme d'enseignement (présentiel ou non). Ces ressources pédagogiques numériques s'adressent autant aux enseignants qu'aux étudiants. Elles s'inscrivent dans les parcours de formation et sont validées par les communautés scientifiques des UNT.

Les ressources mutualisées par les établissements adhérents peuvent être gratuites et ouvertes à tous, gratuites et réservées aux établissements adhérents (formation initiale), payantes c.à.d. commercialisées par les établissements adhérents (ressources intégrées dans des formations continues).

Il faudrait rappeler que les UNT ne se substituent en aucun cas aux établissements eux-mêmes mais elles apportent un complément pédagogique à leur enseignement. Malgré leur nom, les Universités Numériques ne sont pas des opérateurs de formation, ne délivrent pas elles-mêmes des diplômes et n'inscrivent pas d'étudiants. C'est le rôle des établissements d'Universités et écoles membres des UNT. C'est la raison pour laquelle les UNT sont qualifiées d'organismes « sans murs », fédérant des Campus Universitaires installés dans plusieurs universités ou grandes écoles, sur des compétences complémentaires. Elles regroupent non seulement la plupart des Campus numériques français issus des appels à projet lancés par le ministère de l'Éducation nationale, de l'Enseignement supérieur et de la Recherche, mais également des établissements qui se sont engagés ultérieurement dans la conception de contenus numériques.

Il existe huit UNT qui couvrent les domaines des sciences de la santé et du sport, des sciences de l'ingénieur et technologie, de l'économie et gestion, de l'environnement et développement durable, des sciences humaines et sociales, des langues et cultures, des sciences juridiques et politiques, des sciences fondamentales et de l'enseignement technologique.

Pour le besoin de cette étude, les portails des UNT seront analysés à travers l'ensemble des critères définis précédemment (Cf. point 7).

### 8.2.1 AUNEGE (ASSOCIATION DES UNIVERSITES POUR LE DEVELOPPEMENT DE L'ENSEIGNEMENT NUMERIQUE EN ECONOMIE ET GESTION)

AUNEGE est la vitrine des universités françaises en économie et gestion sur le net. Elle définit ses missions dans les termes suivants :

- Cofinancer et développer des ressources pour l'enseignement numérique en s'appuyant sur les compétences humaines et les moyens financiers mutualisés par les établissements membres : (cours complets, fondamentaux, études de cas, jeux et simulations, témoignages et reportages)

- Assurer une veille et la distribuer auprès des enseignants, des personnels et des étudiants de nos établissements sur les politiques, les méthodologies, les normes et les standards en matière d'enseignement numérique,

- Mettre ces ressources à disposition des utilisateurs selon des critères fondés sur des principes de qualité et répondant à une politique de distribution et de diffusion équitable. Ainsi, le Conseil d'Administration du 14 décembre 2010 a adopté le principe de l'accès libre des ressources AUNEGE à tous les publics.

✓ **Catégorisation (données d'identité)**

| | |
|---|---|
| Nom/sigle du portail ; | 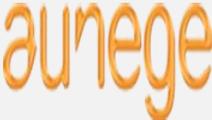<br>http://www.aunege.fr/ |
| Nature du portail (préciser s'il s'agit d'un portail du secteur public, privé ou associatif) ; | Associatif |
| Mode d'accès aux ressources («gratuit» ou «payant» ; «ouvert» ou «sur inscription») ; | Ouvert et gratuit (avec espace réservé aux membres inscrits) |
| Tutelle et/ou porteur du portail | Université Paris Ouest Nanterre La Défense |
| Population cible (distinguer si le portail est pour étudiants et/ou enseignants) ; | Apprenant, enseignant |
| Aspects linguistiques (contenus multilingues) ; | Multilingue |
| Nombre de ressources pédagogiques accessibles ; | Sciences économiques [277]<br><br>Sciences de gestion [406]<br><br>Outils, méthodes et disciplines connexes [90] |
| Niveau académique des ressources disponibles (collège, cycles LMD, formation professionnelle, etc.) ; | LMD, formation professionnelle |

| Typologie informatique des ressources disponibles (fichiers texte, multimédias) | Texte, vidéo, son, animation, collection, ensemble de données, évènement, image, image en mouvement, image fixe, logiciel, objet physique, ressources interactive, service |
|---|---|
| Typologies éducative des ressources pédagogiques (cours, exercices, démos, etc.) ; | exercice, simulation, questionnaire, examen, expérience, autoévaluation, cours, présentation, énoncé de problème, démonstration, animation, tutoriel, glossaire, guide, matériel de référence, méthodologie, outil, scénario pédagogique, évaluation, étude de cas, liste de références, jeu de données, autres |

- ✓ **Organisation**

| Les ressources sont disposées en collections (thématiques, niveaux scolaires, types de supports, etc.) ; | Classification thématique en mode annuaire |
|---|---|
| Le portail dispose de sa propre base de ressources (réservoir d'objets pédagogiques) accessibles par ses propres outils de recherche (annuaire et moteur) [vérification par URL] ; | Oui en partie (ressources propres à AUNEGE) |
| Le portail dispose d'une base de référence (de métadonnées) liée à des ressources hébergées sur des serveurs externes ; | Oui en partie (ressources sur Canal-U) |
| Le portail ne dispose ni bases de références ni de base de ressources ni d'outils de recherche propres mais propose une liste de signets vers des ressources pédagogiques sur des sites externes ; | Non : portail avec un système d'information propre et des services à valeurs ajoutées |
| L'alimentation du portail en ressources pédagogiques est soit réservée à un personnel spécialisé (auteurs et documentalistes) ou ouverte au public (sur validation préalable) ; | Soumission de projets de contenus par tous les établissements du supérieur, prioritairement les membres d'AUNEGE |
| Données sur les droits d'auteurs des | Oui |

| ressources pédagogiques. | |
|---|---|

✓ **Description/moissonnage**

| Résumés (vérifier si les ressources sont accompagnées de résumés synoptiques) | Oui |
|---|---|
| Métadonnées (vérifier si les ressources sont décrites par des métadonnées) | Oui |
| Critères de description pédagogique utilisés (niveau scolaire, durée d'apprentissage, méthode d'apprentissage, type d'évaluation, etc.) | Type pédagogique, Activité induite, Niveau, Public cible, Contexte d'utilisation, Durée d'apprentissage |
| Schéma de métadonnées ou profil d'application (DC, LOM, LOMFR, SupLOMFR, MLR ou format libre) | LOMv1.0, LOMFRv1.0, SupLOMFRv1.0 |
| Vocabulaire ou langage contrôlé (index, listes de choix, taxonomie, thésaurus, etc.) ; | Vocabulaire contrôlé pour les champs Type de contenu, Type de la ressource pédagogique et Niveau |
| Moissonnage OAI-PMH ou autre ; | OAI-PMH |

✓ **Exploitation**

| Mode d'accès aux ressources («gratuit» ou «payant» ; «ouvert» ou «sur inscription») | Ouvert et gratuit |
|---|---|
| Double accès aux ressources en mode annuaire et en mode recherche locale ; | Oui |
| Ressources accessibles par les moteurs de recherche génériques (test Google) ; | Oui |
| Recherche par filtrage multicritère (mode cochage par catégories) ; | Non |
| Recherche booléenne (et, ou, sauf) | Oui |

| Services de veille (alertes, RSS…) | Oui (Rss) sur mode annuaire et resultats de recherche, bookmarking social Pearltrees |
|---|---|
| Outils de partage Web 2.0 (Facebook, Twitter, etc.) | Twitter, Facebook, |
| Affichage automatique de listes de nouveaux contenus ; | Nouvelles ressources |
| Reconnaissance automatique par les outils bibliographiques (Test Zotero) | Non |
| Exportation des références bibliographiques (formats Zotero, CSV, XML, etc.) | XML (fiches de métadonnées) |
| Connecteurs à des applications externes (Moodle, etc.) | Non |

### 8.2.2 UOH (UNIVERSITE OUVERTE POUR LES HUMANITES)

Créée sous l'impulsion du Ministère de l'Enseignement Supérieur et de la Recherche (MESR), l'Université Ouverte des Humanités est l'Université Numérique Thématique consacrée aux champs disciplinaires des Sciences humaines, des Sciences sociales, des Lettres, des Langues et des Arts.

Pour favoriser une meilleure réussite des étudiants, notamment en licence, et contribuer au développement de l'université numérique française, l'UOH offre sur ce portail en libre accès des contenus pédagogiques validés scientifiquement, pédagogiquement et techniquement.

Regroupement d'établissements supérieurs mettant en commun leurs moyens, l'UOH n'est pas une université française, et ne se substitue pas aux établissements supérieurs. Les ressources qu'elle propose sont des compléments et/ou des supports aux cours qui permettent la diversification des modes de transmission des connaissances et offrent la possibilité à tous les établissements du supérieur de construire des stratégies d'enseignement s'ils le désirent.

23 Universités françaises, une Ecole Normale Supérieure, une Université belge, une Université camerounaise et une Université canadienne font aujourd'hui partie du projet. Ce qui représente potentiellement plus de 425 000 étudiants.

Depuis sa fondation en tant que service commun interuniversitaire début 2007, l'UOH œuvre pour la valorisation des ressources pédagogiques numériques des établissements partenaires et publie des appels à projets encourageant la production mutualisée de ressources pédagogiques numériques.

✓ **Catégorisation (données d'identité)**

| Nom/sigle du portail ; | 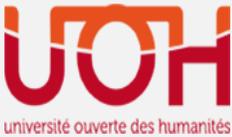<br>http://www.uoh.fr/ |
|---|---|
| Nature du portail (préciser s'il s'agit d'un portail du secteur public, privé ou associatif) ; | Regroupement d'établissements d'enseignement supérieur |
| Mode d'accès aux ressources («gratuit» ou «payant» ; «ouvert» ou «sur inscription») ; | Ouvert et gratuit (avec espace réservé aux membres inscrits) |
| Tutelle et/ou porteur du portail | L'hébergement du serveur de production est assuré par la Direction Informatique de l'Université de Strasbourg |
| Population cible (distinguer si le portail est pour étudiants et/ou enseignants) ; | Apprenants, enseignants |
| Aspects linguistiques (contenus multilingues) ; | Multilingue |
| Nombre de ressources pédagogiques accessibles ; | 1471 Ressources estampillées UOH et Cours complets |
| Niveau académique des ressources disponibles (collège, cycles LMD, formation professionnelle, etc.) ; | Doctorat, Formation continue, Formation en entreprise, Formation professionnelle, LMD |
| Typologie informatique des ressources disponibles (fichiers texte, multimédias) | Image, Logiciel, Ressource interactive, Son, Texte, Vidéo |
| Typologies éducative des ressources pédagogiques (cours, exercices, démos, etc.) ; | Grande Leçon (type UOH), Webographie, bibliographie, Essentiel (type UOH), Exercice, Guide pédagogique, Autoévaluation, Etude et document (type UOH), Glossaire, Web documentaire, Diaporama, Index, Conférence, table ronde et entretien (type UOH), Simulation, Méthodologie, Etude de cas, Jeu sérieux |

- ✓ **Organisation**

| Les ressources sont disposées en collections (thématiques, niveaux scolaires, types de supports, etc.) ; | Disciplines, types pédagogiques, types documentaires, Niveau d'instruction, âge de la ressource |
|---|---|
| Le portail dispose de sa propre base de ressources (réservoir d'objets pédagogiques) accessibles par ses propres outils de recherche (annuaire et moteur) [vérification par URL] ; | Non : ressources hébergées sur serveurs externes. Extrait : « Les établissements partenaires de l'UOH mettent leurs meilleures ressources pédagogiques en ligne sur ce portail. » |
| Le portail dispose d'une base de référence (de métadonnées) qui pointent vers des ressources hébergées sur des serveurs externes ; | Oui (ressources hébergées sur serveurs externes) |
| Le portail ne dispose ni bases de références ni de base de ressources ni d'outils de recherche propres mais propose une liste de signets vers des ressources pédagogiques sur des sites externes ; | Non : portail avec des services à valeurs ajoutées |
| L'alimentation du portail en ressources pédagogiques est soit réservée à un personnel spécialisé (auteurs et documentalistes) ou ouverte au public (sur validation préalable) ; | Réservé à un réseau d'enseignants-chercheurs producteurs parmi les 64 universités dispensant des formations dans le domaine des Humanités. Les commentaires des ressources référencées sont publics de manière anonyme ou identifiée |
| Données sur les droits d'auteurs des ressources pédagogiques. | Oui (essentiellement creativeCommons) |

- ✓ **Description/moissonnage**

| Résumés (vérifier si les ressources sont accompagnées de résumés synoptiques) | Oui |
|---|---|
| Métadonnées (vérifier si les ressources sont décrites par des métadonnées) | Oui |
| Critères de description pédagogique utilisés (niveau scolaire, durée d'apprentissage, méthode | Type pédagogique, Niveau, Activité induite, Proposition d'utilisation, Durée d'apprentissage, |
| d'apprentissage, type d'évaluation, etc.) | Langue de l'apprenant |
| Schéma de métadonnées ou profil d'application (DC, LOM, LOMFR, SupLOMFR, MLR ou format libre) | NA |
| Vocabulaire ou langage contrôlé (index, listes de choix, taxonomie, thésaurus, etc.) ; | Listes de choix (sur recherche avancée), liste autorités Auteurs, Rameau, Vocabulaire libre |
| Moissonnage OAI-PMH ou autre ; | NA |

- ✓ **Exploitation**

| | |
|---|---|
| Mode d'accès aux ressources («gratuit» ou «payant» ; «ouvert» ou «sur inscription») | Gratuit & public |
| Double accès aux ressources en mode annuaire et en mode recherche locale ; | Oui |
| Ressources accessibles par les moteurs de recherche génériques (test Google) ; | Oui |
| Recherche par filtrage multicritère (mode cochage par catégories) ; | Oui (sélection par marquage de critères) |
| Recherche booléenne (et, ou, sauf) | Non |
| Services de veille (alertes, RSS…) | |
| Outils de partage Web 2.0 (Facebook, Twitter, etc.) | Oui (FB, Twitter, Linkedin) |
| Affichage automatique de listes de nouveaux contenus ; | Oui (en vignette) |
| Reconnaissance automatique par les outils bibliographiques (Test Zotero) | Non |
| Exportation des références bibliographiques (formats Zotero, CSV, XML, etc.) | XML (Suplomfr) |

| Connecteurs à des applications externes (Moodle, etc.) | Isidore et Worldcat |

## 8.2.3 UNISCIEL (UNIVERSITE DES SCIENCES EN LIGNE)

Unisciel est un GIS (Groupement d'intérêt Scientifique). Il est l'une des 7 UNT (Université Numérique Thématique) françaises. Unisciel met à disposition de tous plus de 4000 ressources pédagogiques en Mathématiques, Physique, Chimie, Informatique, Sciences de la Vie, de la Terre et de l'Univers, sous forme de cours, exercices, vidéos, animations, tests... Plusieurs dispositifs ont également été créés afin d'aider les étudiants, les enseignants et les établissements dans leurs formations et leurs missions.

✓ **Catégorisation (données d'identité)**

| | |
|---|---|
| Nom/sigle du portail ; | 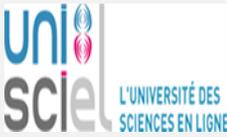 http://www.unisciel.fr/ |
| Nature du portail (préciser s'il s'agit d'un portail du secteur public, privé ou associatif) ; | Associatif (groupement d'intérêt scientifique) |
| Mode d'accès aux ressources («gratuit» ou «payant» ; «ouvert» ou «sur inscription») ; | Ouvert et gratuit (avec espace réservé aux membres inscrits) |
| Tutelle et/ou porteur du portail | Université des Sciences et Technologies de Lille – Lille 1 |
| Population cible (distinguer si le portail est pour étudiants et/ou enseignants) ; | Collégiens, lycéens, étudiants, enseignants |
| Aspects linguistiques (contenus multilingues) ; | Multilingue |
| Nombre de ressources pédagogiques accessibles ; | 4411 ressources de type modules ou chapitres |
| Niveau académique des ressources disponibles (collège, cycles LMD, formation professionnelle, etc.) ; | Majorité terminale et licence |
| Typologie informatique des ressources | Liens, documents téléchargeables, ressources |

| disponibles (fichiers texte, multimédias) | interactives, textes, vidéos, images, dossiers thématiques, Podcast audio, Logiciels |
|---|---|
| Typologies éducative des ressources pédagogiques (cours, exercices, démos, etc.) ; | Cours, exercices corrigés, annales, fiches pratiques, QCM, études de cas, expériences-TP, animations interactives |

- ✓ **Organisation**

| Les ressources sont disposées en collections (thématiques, niveaux scolaires, types de supports, etc.) ; | Listes thématiques |
|---|---|
| Le portail dispose de sa propre base de ressources (réservoir d'objets pédagogiques) accessibles par ses propres outils de recherche (annuaire et moteur) [vérification par URL] ; | Oui en partie (ressources UNISCIEL) |
| Le portail dispose d'une base de référence (de métadonnées) qui pointent vers des ressources hébergées sur des serveurs externes ; | Oui en partie (ressources sur sites externes) |
| Le portail ne dispose ni bases de références ni de base de ressources ni d'outils de recherche propres mais propose une liste de signets vers des ressources pédagogiques sur des sites externes ; | Non : portail avec des services à valeurs ajoutées |
| L'alimentation du portail en ressources pédagogiques est soit réservée à un personnel spécialisé (auteurs et documentalistes) ou ouverte au public (sur validation préalable) ; | réservées |
| Données sur les droits d'auteurs des ressources pédagogiques. | Oui (GNU GPL ; Creativecommons) |

- ✓ **Description/moissonnage**

| Résumés (vérifier si les ressources sont accompagnées de résumés | Oui |
|---|---|

| synoptiques) | |
|---|---|
| Métadonnées (vérifier si les ressources sont décrites par des métadonnées) | Tags |
| Critères de description pédagogique utilisés (niveau scolaire, durée d'apprentissage, méthode d'apprentissage, type d'évaluation, etc.) | Niveau scolaire |
| Schéma de métadonnées ou profil d'application (DC, LOM, LOMFR, SupLOMFR, MLR ou format libre) | NA |
| Vocabulaire ou langage contrôlé (index, listes de choix, taxonomie, thésaurus, etc.) ; | Tags en langage libre |
| Moissonnage OAI-PMH ou autre ; | OAI-PMH |

- ✓ **Exploitation**

| Double accès aux ressources en mode annuaire et en mode recherche locale ; | Oui |
|---|---|
| Ressources accessibles par les moteurs de recherche génériques (test Google) ; | Oui |
| Recherche par filtrage multicritère (mode cochage par catégories) ; | Oui (par filtres) |
| Recherche booléenne (et, ou, sauf) | Non |
| Services de veille (alertes, RSS…) | Non |
| Outils de partage Web 2.0 (Facebook, Twitter, etc.) | Twitter, FB, G+ |
| Affichage automatique de listes de nouveaux contenus ; | Non |
| Reconnaissance automatique par les outils bibliographiques (Test Zotero) | Non |

| Exportation des références bibliographiques (formats Zotero, CSV, XML, etc.) | Non |
|---|---|
| Connecteurs à des applications externes (Moodle, etc.) | Projets ressources spécifiques de culture scientifique :<br>- Le Mooc Quidquam?,<br>- CAPA Chaine éditorial,<br>- BEEBAC moteur de recherche et réseau social |

## 8.2.4 UNIT (UNIVERSITE NUMERIQUE, INGENIERIE ET TECHNOLOGIE)

UNIT, l'Université Numérique Ingénierie et Technologie, est l'une des Universités Numériques Thématiques nationales (UNT) créées à l'initiative de Grandes Écoles, d'Universités et du Ministère chargé de l'Enseignement Supérieur. Elle associe tous les acteurs publics et privés de la formation supérieure en Sciences de l'Ingénieur et Technologie désireux de partager des documents numériques existants, des outils, des expériences et de co-piloter des projets basés sur les TICE.

Le portail ORI-OAI d'UNIT est au service d'une politique d'aide à la production et à la mutualisation de ressources pédagogiques numériques. L'objectif est de donner à tous les enseignants et étudiants en sciences de l'ingénieur et en ingénierie, le libre accès à un ensemble croissant d'enseignements en visant une large diffusion sur Internet de ressources numériques de qualité. L'ambition est également de renforcer la visibilité des formations offertes par les membres d'UNIT, en s'appuyant pour cela sur le potentiel d'échange et de diffusion offert par le réseau de portails OAI.

UNIT propose aujourd'hui, en libre accès, une offre documentaire de près de 2500 ressources pédagogiques. Ce catalogue interactif couvre tous les grands domaines disciplinaires des Sciences de l'Ingénieur et de la Technologie : ingénierie de l'environnement, énergétique, mécanique, science des matériaux, génie des procédés, automatique, électronique, électricité et électrotechnique, modélisation et simulation, informatique, télécommunications, optique, etc.

À la constitution de sa bibliothèque de contenus pédagogiques numérisés, UNIT avait le choix entre un serveur central classique ou une architecture distribuée gérant les ressources au plus près de leurs auteurs. Le choix de ce second schéma a engendré de nombreux avantages, notamment la mutualisation et l'interopérabilité des ressources produites par chaque établissement et par chaque auteur. Une architecture distribuée permettrait de répondre aux objectifs suivants :

- La mutualisation des ressources pédagogiques des établissements partenaires afin d'accroître leur diffusion et leur utilisation, à la fois par les enseignants et par les étudiants ;

- La simplification de l'accès aux documents en local ou à distance de façon identique et transparente ;

- La gestion fine des droits d'accès selon les besoins des établissements concernés ;

- Une meilleure gestion des doublons en limitant l'usage des documents à une seule version, ce qui permet de garantir leur mise à jour, d'alléger la charge informatique et réseau en évitant de copier de multiples fois les mêmes documents.

✓ **Catégorisation (données d'identité)**

| | |
|---|---|
| Nom/sigle du portail ; | 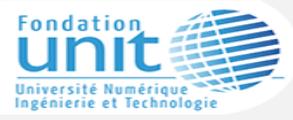 http://www.unit.eu/fr |
| Nature du portail (préciser s'il s'agit d'un portail du secteur public, privé ou associatif) ; | Associatif (fondation), acteurs publics et privés |
| Mode d'accès aux ressources («gratuit» ou «payant» ; «ouvert» ou «sur inscription») ; | Ouvert et gratuit (avec espace réservé aux membres inscrits) |
| Tutelle et/ou porteur du portail | NA |
| Population cible (distinguer si le portail est pour étudiants et/ou enseignants) ; | élèves ingénieurs, étudiants de licences professionnelles, IUT, BTS, techniciens et cadres techniques des entreprises, grand public |
| Aspects linguistiques (contenus multilingues) ; | Multilingue |
| Nombre de ressources pédagogiques accessibles ; | 2500 |
| Niveau académique des ressources disponibles (collège, cycles LMD, formation professionnelle, etc.) ; | Scolaire (primaire, secondaire), Supérieur (LMD), Professionnel |
| Typologie informatique des ressources disponibles (fichiers texte, multimédias) | Document HTML, Document Flash |
| Typologies éducative des ressources pédagogiques (cours, exercices, démos, etc.) ; | Grain, leçon, cours, module, exercice, examen, simulation, questionnaire, expérience, autoévaluation, cours/présentation, démonstration, animation, tutoriel, glossaire, guide, matériel de référence, méthodologie, scénario pédagogique, études de cas, |

| | références, jeux de données, |
|---|---|

- ✓ **Organisation**

| | |
|---|---|
| Les ressources sont disposées en collections (thématiques, niveaux scolaires, types de supports, etc.) ; | Annuaire thématique, par établissement source, par auteur |
| Le portail dispose de sa propre base de ressources (réservoir d'objets pédagogiques) accessibles par ses propres outils de recherche (annuaire et moteur) [vérification par URL] ; | Non |
| Le portail dispose d'une base de référence (de métadonnées) qui pointent vers des ressources hébergées sur des serveurs externes ; | Oui |
| Le portail ne dispose ni bases de références ni de base de ressources ni d'outils de recherche propres mais propose une liste de signets vers des ressources pédagogiques sur des sites externes ; | Non : portail avec des services à valeurs ajoutées |
| L'alimentation du portail en ressources pédagogiques est soit réservée à un personnel spécialisé (auteurs et documentalistes) ou ouverte au public (sur validation préalable) ; | Réservé |
| Données sur les droits d'auteurs des ressources pédagogiques. | Oui (Creative Commons) |

- ✓ **Description/moissonnage**

| | |
|---|---|
| Résumés (vérifier si les ressources sont accompagnées de résumés synoptiques) | Oui |
| Métadonnées (vérifier si les ressources sont décrites par des métadonnées) | Oui |

| Critères de description pédagogique utilisés (niveau scolaire, durée d'apprentissage, méthode d'apprentissage, type d'évaluation, etc.) | Type pédagogique, Granularité, Niveau, Age attendu du l'utilisateur, Public cible, Langue de l'apprenant, Proposition d'utilisation, Durée d'apprentissage |
|---|---|
| Schéma de métadonnées ou profil d'application (DC, LOM, LOMFR, SupLOMFR, MLR ou format libre) | LOMv1.0, LOMFRv1.0, SupLOMFRv1.0 |
| Vocabulaire ou langage contrôlé (index, listes de choix, taxonomie, thésaurus, etc.) ; | Mots clés libres |
| Moissonnage OAI-PMH ou autre ; | Réseau de portails basé sur le logiciel libre : ORI-OAI |

- ✓ **Exploitation**

| Double accès aux ressources en mode annuaire et en mode recherche locale ; | Oui |
|---|---|
| Ressources accessibles par les moteurs de recherche génériques (test Google) ; | Oui |
| Recherche par filtrage multicritère (mode cochage par catégories) ; | Oui |
| Recherche booléenne (et, ou, sauf) | Oui |
| Services de veille (alertes, RSS…) | |
| Outils de partage Web 2.0 (Facebook, Twitter, etc.) | Non |
| Affichage automatique de listes de nouveaux contenus ; | Rubrique « A découvrir » et « dernières mises en ligne » |
| Reconnaissance automatique par les outils bibliographiques (Test Zotero) | Non |
| Exportation des références bibliographiques (formats Zotero, CSV, XML, etc.) | XML (fiche de description) |

| Connecteurs à des applications externes (Moodle, etc.) | Non |
|---|---|

## 8.2.5 UNJF (UNIVERSITE NUMERIQUE JURIDIQUE FRANCOPHONE) -

L'Université Numérique Juridique Francophone (UNJF) est une des sept « Universités Numériques Thématiques » nées d'un processus de mutualisation des ressources pédagogiques universitaires soutenue par l'Etat (Ministère de l'Enseignement Supérieur et de la Recherche).

Elle se développe en relation avec la Conférence des Doyens des Facultés de Droit qui est l'instance la plus représentative des responsables de la formation juridique en milieu universitaire, selon un processus de mutualisation voulu par les universités françaises et appuyé par la Conférence des Présidents d'Universités.

Sa principale mission est de répondre aux besoins de la formation à distance dans le domaine juridique en proposant une offre complète et diversifiée.

L'UNJF a ainsi vocation à devenir le centre de ressources de référence des universités, Facultés et centres de formation et d'enseignement qui souhaitent proposer des Formations Ouvertes et à Distance (FOAD) dans les disciplines juridiques.

✓ **Catégorisation (données d'identité)**

| Nom/sigle du portail ; | 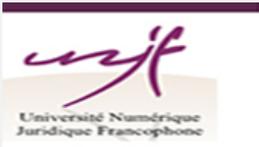 http://www.unjf.fr/ |
|---|---|
| Nature du portail (préciser s'il s'agit d'un portail du secteur public, privé ou associatif) ; | Associatif |
| Mode d'accès aux ressources («gratuit» ou «payant» ; «ouvert» ou «sur inscription») ; | Gratuit mais sur inscription (espace réservé aux membres inscrits) |
| Tutelle et/ou porteur du portail | Centre national d'enseignement à distance, Université de Paris 1 Panthéon-Sorbonne |
| Population cible (distinguer si le portail | Apprenants, enseignants |

| est pour étudiants et/ou enseignants) ; | |
|---|---|
| Aspects linguistiques (contenus multilingues) ; | Multilingue |
| Nombre de ressources pédagogiques accessibles ; | NA |
| Niveau académique des ressources disponibles (collège, cycles LMD, formation professionnelle, etc.) ; | Licence, Master |
| Typologie informatique des ressources disponibles (fichiers texte, multimédias) | NA |
| Typologies éducative des ressources pédagogiques (cours, exercices, démos, etc.) ; | Cours, actualités, formations |

✓ **Organisation**

| Les ressources sont disposées en collections (thématiques, niveaux scolaires, types de supports, etc.) ; | Par institutions membres, par niveaux de Licence et Master |
|---|---|
| Le portail dispose de sa propre base de ressources (réservoir d'objets pédagogiques) accessibles par ses propres outils de recherche (annuaire et moteur) [vérification par URL] ; | Non (centre de ressources de référence pour les universités) |
| Le portail dispose d'une base de référence (de métadonnées) qui pointent vers des ressources hébergées sur des serveurs externes ; | Oui (portail de services) |
| Le portail ne dispose ni bases de références ni de base de ressources ni d'outils de recherche propres mais propose une liste de signets vers des ressources pédagogiques sur des sites externes ; | Non : portail avec des services à valeurs ajoutées |
| L'alimentation du portail en ressources pédagogiques est soit réservée à un | S'effectue au niveau des institutions membres |

| personnel spécialisé (auteurs et documentalistes) ou ouverte au public (sur validation préalable) ; | |
| --- | --- |
| Données sur les droits d'auteurs des ressources pédagogiques. | Non |

- ✓ **Description/moissonnage**

| Résumés (vérifier si les ressources sont accompagnées de résumés synoptiques) | Oui |
| --- | --- |
| Métadonnées (vérifier si les ressources sont décrites par des métadonnées) | Oui (libres) |
| Critères de description pédagogique utilisés (niveau scolaire, durée d'apprentissage, méthode d'apprentissage, type d'évaluation, etc.) | Niveau |
| Schéma de métadonnées ou profil d'application (DC, LOM, LOMFR, SupLOMFR, MLR ou format libre) | NA |
| Vocabulaire ou langage contrôlé (index, listes de choix, taxonomie, thésaurus, etc.) ; | Vocabulaire libre |
| Moissonnage OAI-PMH ou autre ; | |

- ✓ **Exploitation**

| Double accès aux ressources en mode annuaire et en mode recherche locale ; | Oui |
| --- | --- |
| Ressources accessibles par les moteurs de recherche génériques (test Google) ; | Non (restriction d'accès pour les catalogues de bibliothèques membres) |
| Recherche par filtrage multicritère (mode cochage par catégories) ; | Non |
| Recherche booléenne (et, ou, sauf) | Non |
| Services de veille (alertes, RSS…) | Non |
| Outils de partage Web 2.0 (Facebook, Twitter, etc.) | FB, Twitter |
| Affichage automatique de listes de nouveaux contenus ; | Oui (cours en démo) |
| Reconnaissance automatique par les outils bibliographiques (Test Zotero) | Non |
| Exportation des références bibliographiques (formats Zotero, CSV, XML, etc.) | Non |
| Connecteurs à des applications externes (Moodle, etc.) | Nomodos, revue d'histoire |

## 8.2.6 UVED (UNIVERSITE VIRTUELLE ENVIRONNEMENT ET DEVELOPPEMENT DURABLE)

Créée en juin 2005, l'Université Virtuelle Environnement et Développement durable (UVED) est l'une des sept Universités Numériques Thématiques (UNT) soutenues par le Ministère de l'Enseignement supérieur et de la Recherche.

Après plus de cinq années d'activité sous la forme associative, la forme juridique d'UVED a évolué en fondation partenariale pour pouvoir élargir et renforcer ses activités, notamment en associant des établissements publics et privés. L'arrêté d'autorisation de création de la Fondation partenariale UVED a été publié au Bulletin Officiel n° 28 du 14 juillet 2011 du Ministère de l'Enseignement supérieur et de la Recherche marquant ainsi l'acte de naissance de la Fondation.

L'UVED œuvre à la constitution et à la valorisation nationale et internationale du patrimoine pédagogique numérique français validé d'un point de vue scientifique, pédagogique et technique et au rayonnement de l'enseignement supérieur français au sein du monde francophone.

Chaque ressource pédagogique numérique produite sur financement UVED comporte une partie 'cours' ou 'contenu' et une partie 'kit pédagogique', un guide d'usage et d'accompagnement de la ressource qui définit les activités liées à la ressource et les méthodes d'utilisation (contextes, parcours possibles...) permettant la mise en place et le bon déroulement de la formation des apprenants.

✓ **Catégorisation (données d'identité)**

| Nom/sigle du portail ; | 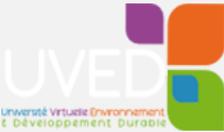<br>http://www.uved.fr/navigation/accueil.html |
|---|---|
| Nature du portail (préciser s'il s'agit d'un portail du secteur public, privé ou associatif) ; | Associatif (fondation) |
| Mode d'accès aux ressources («gratuit» ou «payant» ; «ouvert» ou «sur inscription») ; | Ouvert et gratuit (avec espace réservé aux membres inscrits) |
| Tutelle et/ou porteur du portail | NA |
| Population cible (distinguer si le portail est pour étudiants et/ou enseignants) ; | Apprenant, enseignant, gestionnaire, professionnel |
| Aspects linguistiques (contenus multilingues) ; | Multilingue |
| Nombre de ressources pédagogiques accessibles ; | + 300 |
| Niveau académique des ressources disponibles (collège, cycles LMD, formation professionnelle, etc.) ; | Enseignement scolaire, Enseignement secondaire, Enseignement supérieur (LMD), professionnel, formation continue, formation en entreprise |
| Typologie informatique des ressources disponibles (fichiers texte, multimédias) | Web |
| Typologies éducative des ressources pédagogiques (cours, exercices, démos, etc.) ; | Autoévaluation, cours, présentations, études de cas, examen, exercice, glossaire, guide, méthodologie, outil, questionnaire, expérience |

✓ **Organisation**

| Les ressources sont disposées en collections (thématiques, niveaux scolaires, types de supports, etc.) ; | Liste par thèmes (mode cartographique), liste thématique alphabétique (pédagothèque) |
|---|---|
| Le portail dispose de sa propre base de ressources (réservoir d'objets | Non |

| pédagogiques) accessibles par ses propres outils de recherche (annuaire et moteur) [vérification par URL] ; | |
|---|---|
| Le portail dispose d'une base de référence (de métadonnées) qui pointent vers des ressources hébergées sur des serveurs externes ; | Oui « UVED recense, référence et valorise les ressources pédagogiques numériques existantes au sein de ses établissements partenaires » |
| Le portail ne dispose ni bases de références ni de base de ressources ni d'outils de recherche propres mais propose une liste de signets vers des ressources pédagogiques sur des sites externes ; | Non : portail avec des services à valeurs ajoutées |
| L'alimentation du portail en ressources pédagogiques est soit réservée à un personnel spécialisé (auteurs et documentalistes) ou ouverte au public (sur validation préalable) ; | Réservée aux établissements partenaires |
| Données sur les droits d'auteurs des ressources pédagogiques. | Non |

✓ **Description/moissonnage**

| Résumés (vérifier si les ressources sont accompagnées de résumés synoptiques) | Oui |
|---|---|
| Métadonnées (vérifier si les ressources sont décrites par des métadonnées) | Oui |
| Critères de description pédagogique utilisés (niveau scolaire, durée d'apprentissage, méthode d'apprentissage, type d'évaluation, etc.) | Type de ressource, public cible, niveau d'apprentissage, âge d'apprenant, durée d'apprentissage, langue |
| Schéma de métadonnées ou profil d'application (DC, LOM, LOMFR, SupLOMFR, MLR ou format libre) | LOMv1.0, LOMFRv1.0, SupLOMFRv1.0 |

| Vocabulaire ou langage contrôlé (index, listes de choix, taxonomie, thésaurus, etc.) ; | Vocabulaire libre (descripteurs) |
|---|---|
| Moissonnage OAI-PMH ou autre ; | |

- ✓ **Exploitation**

| Double accès aux ressources en mode annuaire et en mode recherche locale ; | Oui |
|---|---|
| Ressources accessibles par les moteurs de recherche génériques (test Google) ; | Oui |
| Recherche par filtrage multicritère (mode cochage par catégories) ; | Oui |
| Recherche booléenne (et, ou, sauf) | Non |
| Services de veille (alertes, RSS…) | RSS (actualité) |
| Outils de partage Web 2.0 (Facebook, Twitter, etc.) | Non |
| Affichage automatique de listes de nouveaux contenus ; | Rubrique « dernières ressources » |
| Reconnaissance automatique par les outils bibliographiques (Test Zotero) | Non |
| Exportation des références bibliographiques (formats Zotero, CSV, XML, etc.) | XML (résultat de recherche) |
| Connecteurs à des applications externes (Moodle, etc.) | Non |

## 8.2.7 UNF3S (UNIVERSITE NUMERIQUE FRANCOPHONE DES SCIENCES DE LA SANTE ET DU SPORT)

L'UNF3S a été créée en 2003 pour la médecine sous le nom de l'UMVF (Université médicale virtuelle francophone). En 2009, elle devient UNF3S en intégrant les disciplines pharmacie, odontologie et sciences du sport. Il existe sept Universités Numériques Thématiques de l'enseignement supérieur.

Trente-sept universités françaises ayant des composantes en médecine, pharmacie, odontologie ou sport adhèrent à ce groupement d'intérêt public, cofinancé par ces universités elles-mêmes et par des subventions du Ministère de l'Enseignement Supérieur et de la Recherche.

L'UNF3S propose, via ses sites internet, en accès libre et gratuit, des ressources pédagogiques classiques (pdf, qcm, cours ou colloques scientifiques filmés) et innovantes (qcm interactifs, diaporamas sonorisés et chapitrés, représentations en 3D, jeux sérieux, simulations) validées par les collèges de spécialités de ses composantes ou issues des sites des facultés françaises, par spécialité ou par année d'étude.

✓ **Catégorisation (données d'identité)**

| | |
|---|---|
| Nom/sigle du portail ; | 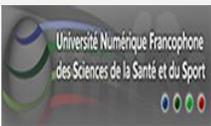<br>http://www.unf3s.org/ |
| Nature du portail (préciser s'il s'agit d'un portail du secteur public, privé ou associatif) ; | Associatif (4 composantes thématiques) |
| Mode d'accès aux ressources («gratuit» ou «payant» ; «ouvert» ou «sur inscription») ; | Ouvert et gratuit (avec espace réservé aux membres inscrits) |
| Tutelle et/ou porteur du portail | Faculté de médecine, Lille (siège social) |
| Population cible (distinguer si le portail est pour étudiants et/ou enseignants) ; | Apprenants, enseignants, auteurs, gestionnaire |
| Aspects linguistiques (contenus multilingues) ; | Multilingue |
| Nombre de ressources pédagogiques accessibles ; | NA |
| Niveau académique des ressources disponibles (collège, cycles LMD, | NA |

| formation professionnelle, etc.) ; | |
|---|---|
| Typologie informatique des ressources disponibles (fichiers texte, multimédias) | (Sur sites partenaires) : Sites Web, Images, Vidéo, médias enrichis, pdf, représentations en 3D, |
| Typologies éducative des ressources pédagogiques (cours, exercices, démos, etc.) ; | qcm, podcast, qcm interactifs, diaporamas sonorisés et chapitrés, jeux sérieux, simulations |

- ✓ **Organisation**

| Les ressources sont disposées en collections (thématiques, niveaux scolaires, types de supports, etc.) ; | 4 pointeurs vers les sites thématiques |
|---|---|
| Le portail dispose de sa propre base de ressources (réservoir d'objets pédagogiques) accessibles par ses propres outils de recherche (annuaire et moteur) [vérification par URL] ; | Non (exploite les services et ressources des sites partenaires |
| Le portail dispose d'une base de référence (de métadonnées) qui pointent vers des ressources hébergées sur des serveurs externes ; | Non (juste signets de pointages vers les sites patenaires) |
| Le portail ne dispose ni bases de références ni de base de ressources ni d'outils de recherche propres mais propose une liste de signets vers des ressources pédagogiques sur des sites externes ; | Oui : portail ne fournit aucun service à valeurs ajoutées |
| L'alimentation du portail en ressources pédagogiques est soit réservée à un personnel spécialisé (auteurs et documentalistes) ou ouverte au public (sur validation préalable) ; | Réservée au niveau des établissements partenaires |
| Données sur les droits d'auteurs des ressources pédagogiques. | Non |

✓ **Description/moissonnage**

| Résumés (vérifier si les ressources sont accompagnées de résumés synoptiques) | Oui (sur les sites partenaires) |
|---|---|
| Métadonnées (vérifier si les ressources sont décrites par des métadonnées) | Oui (sur les sites partenaires) |
| Critères de description pédagogique utilisés (niveau scolaire, durée d'apprentissage, méthode d'apprentissage, type d'évaluation, etc.) | Oui (sur les sites partenaires) |
| Schéma de métadonnées ou profil d'application (DC, LOM, LOMFR, SupLOMFR, MLR ou format libre) | Format libre |
| Vocabulaire ou langage contrôlé (index, listes de choix, taxonomie, thésaurus, etc.) ; | Oui (sur les sites partenaires) |
| Moissonnage OAI-PMH ou autre ; | NA |

✓ **Exploitation**

| Double accès aux ressources en mode annuaire et en mode recherche locale ; | Oui (sur les sites partenaires) |
|---|---|
| Ressources accessibles par les moteurs de recherche génériques (test Google) ; | Oui |
| Recherche par filtrage multicritère (mode cochage par catégories) ; | Non |
| Recherche booléenne (et, ou, sauf) | Oui (sur site agriculture) |
| Services de veille (alertes, RSS…) | Non |
| Outils de partage Web 2.0 (Facebook, Twitter, etc.) | Non |
| Affichage automatique de listes de | « Nouveaux cours » |

| nouveaux contenus ; | |
|---|---|
| Reconnaissance automatique par les outils bibliographiques (Test Zotero) | Non |
| Exportation des références bibliographiques (formats Zotero, CSV, XML, etc.) | Non |
| Connecteurs à des applications externes (Moodle, etc.) | Non |

## 8.2.8 IUTENLIGNE (CATALOGUE DE RESSOURCES PEDAGOGIQUES DE L'ENSEIGNEMENT TECHNOLOGIQUE UNIVERSITAIRE)

Le campus numérique IUTenligne est un projet fédérateur né en réponse aux appels à projets « Campus Numérique » lancés par le Ministère de l'Education Nationale en 2000 puis 2002. Il est coordonné par l'Association des Directeurs d'IUT (AssoDIUT).

La mission d'IUTenligne s'est concrétisée par la création de ressources numériques basées sur les Programmes Pédagogiques Nationaux des Instituts Universitaires Technologiques français dès 2000.

IUTenligne offre des services d'ingénierie pédagogique aux enseignants et enseignants-chercheurs en accompagnant les formateurs d'IUT dans l'utilisation et/ou la création de ressources pédagogiques numériques innovantes.

La médiathèque numérique d'IUTenligne s'adresse aussi bien aux étudiants qu'aux enseignants, qui peuvent puiser les éléments afin d'enrichir leur démarche. Elle dispose d'un ensemble de ressources pédagogiques en accès libre et gratuit en plusieurs formes et formats : Unités d'apprentissage, Cours, Travaux dirigés, Textes, Images, Sons, Exercices et simulateurs interactifs, Travaux pratiques en réseau, Tests d'auto-évaluation, QCM.

✓ **Catégorisation (données d'identité)**

| Nom/sigle du portail ; | 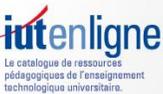<br><br>http://www.iutenligne.net/ |
|---|---|
| Nature du portail (préciser s'il s'agit d'un portail du secteur public, privé ou | Associatif (Association d'IUT) |
| associatif) ; | |
| Mode d'accès aux ressources («gratuit» ou «payant» ; «ouvert» ou «sur inscription») ; | Ouvert et gratuit (avec espace réservé aux membres inscrits) |
| Tutelle et/ou porteur du portail | Association des Directeurs d'IUT (l'ASSODIUT) |
| Population cible (distinguer si le portail est pour étudiants et/ou enseignants) ; | Apprenants, enseignants, auteurs |
| Aspects linguistiques (contenus multilingues) ; | Multilingue |
| Nombre de ressources pédagogiques accessibles ; | 1606 (testé par recherche sur n° de ressource) |
| Niveau académique des ressources disponibles (collège, cycles LMD, formation professionnelle, etc.) ; | LM, formation professionnelle |
| Typologie informatique des ressources disponibles (fichiers texte, multimédias) | Textes, Images, Sons, outils |
| Typologies éducative des ressources pédagogiques (cours, exercices, démos, etc.) ; | Unités d'apprentissage, Cours, Travaux dirigés, Exercices et simulateurs interactifs, Travaux pratiques en réseau, Tests d'auto-évaluation, QCM, Documentation, Etude de cas, Expérience, Méthodologie, Scénario pédagogique, Tutoriels |

✓ **Organisation**

| | |
|---|---|
| Les ressources sont disposées en collections (thématiques, niveaux scolaires, types de supports, etc.) ; | Par catalogue (général, par spécialité, de Quiz) |
| Le portail dispose de sa propre base de ressources (réservoir d'objets pédagogiques) accessibles par ses propres outils de recherche (annuaire et moteur) [vérification par URL] ; | Oui (Ressources propres IUTenligne accessibles par annuaire et moteur de recherche) |
| Le portail dispose d'une base de référence (de métadonnées) qui | Non (base de références et base de documents |

| pointent vers des ressources hébergées sur des serveurs externes ; | propres) |
|---|---|
| Le portail ne dispose ni bases de références ni de base de ressources ni d'outils de recherche propres mais propose une liste de signets vers des ressources pédagogiques sur des sites externes ; | Non : portail avec un système d'information propre et des services à valeurs ajoutées |
| L'alimentation du portail en ressources pédagogiques est soit réservée à un personnel spécialisé (auteurs et documentalistes) ou ouverte au public (sur validation préalable) ; | Réservé aux enseignants universitaires et chargés d'enseignement sur expertise par leurs pairs (médiateurs et experts) pour valider et mettre en forme les propositions de contenus |
| Données sur les droits d'auteurs des ressources pédagogiques. | Oui (creative Commons) |

✓ **Description/moissonnage**

| Résumés (vérifier si les ressources sont accompagnées de résumés synoptiques) | Oui |
|---|---|
| Métadonnées (vérifier si les ressources sont décrites par des métadonnées) | Oui |
| Critères de description pédagogique utilisés (niveau scolaire, durée d'apprentissage, méthode d'apprentissage, type d'évaluation, etc.) | Niveau, Format, Public cible, Approche pédagogique, Approche professionnelle, Méthode d'enseignement, Type de connaissances |
| Schéma de métadonnées ou profil d'application (DC, LOM, LOMFR, SupLOMFR, MLR ou format libre) | NA |
| Vocabulaire ou langage contrôlé (index, listes de choix, taxonomie, thésaurus, etc.) ; | Mots clés libres, listes de choix par diplômes et domaines |
| Moissonnage OAI-PMH ou autre ; | NA |

- ✓ **Exploitation**

| Double accès aux ressources en mode annuaire et en mode recherche locale ; | Oui |
|---|---|
| Ressources accessibles par les moteurs de recherche génériques (test Google) ; | Oui |
| Recherche par filtrage multicritère (mode cochage par catégories) ; | Oui |
| Recherche booléenne (et, ou, sauf) | Oui (par filtrage) |
| Services de veille (alertes, RSS…) | Non |
| Outils de partage Web 2.0 (Facebook, Twitter, etc.) | Oui (FB, Twitter, G+) |
| Affichage automatique de listes de nouveaux contenus ; | « Nouveaux cours » |
| Reconnaissance automatique par les outils bibliographiques (Test Zotero) | Non |
| Exportation des références bibliographiques (formats Zotero, CSV, XML, etc.) | Non |
| Connecteurs à des applications externes (Moodle, etc.) | Plugins, Outils auteurs (Opale, eXeLearning, ELearningMaker, Modulest) |

De l'analyse des portails de ressources pédagogiques déployés par sup-numerique.gouv.fr et notamment les UNT, se dégagent quatre types de configurations :

1. **Portails hybrides** proposant ses propres ressources et des ressources liées (à partir de serveurs d'institutions partenaires) par un système de références locales (métadonnées) qui donne accès (via annuaires et moteurs) à des ressources à la fois en local et sur serveurs externes. C'est le cas des portails AUNEGE, UNISCIEL et FUN-MOOC.

2. **Portails de références** disposant uniquement de bases de références (métadonnées) et d'outils de recherche locale (annuaires et moteurs) donnant accès à des ressources pédagogiques hébergées sur des serveurs d'institutions partenaires. C'est le cas des portails UTO, UNIT, UVED et UNJF.

3. **Portails vitrines** servant uniquement de passerelle vers des serveurs d'institutions partenaires sur lesquels les ressources, les références (métadonnées) et les services de recherche (annuaires et moteurs sont installés). C'est le cas exceptionnel du portail UNF3S.

4. **Portails autonomes** disposant exclusivement de base de ressources propres (contenus pédagogiques des institutions membres) et de base de références propre (métaonnées) et dystsème d'information propre (outils de recherche). C'est l'exemple unique du portail de l'IUTenligne.

Dans ces quatre cas de figures, les portails combinent (à proportions irrégulières) les fonctions d'agrégateur de ressources, de moisonneur de métadonnées, d'interface de recherche ou se présentent comme simple intégrateur de signets. Ils dénotent ainsi qu'un portail de ressources pédagogiques peut être à la fois un fournisseur de données et/ou un fournisseurs de services dans le sens défini plus haut (cf. commentaire sous tableau du portail sup-numerique.gouv.fr). Une configuration ou une autre dépend de plusieurs critères endogènes et exogènes aux parties concernées de la mise en œuvre d'un portail de ressources pédagogiques et devrait faire l'objet d'une étude de faisabilité tenant compte de critères techniques, économiques et ressources humaines.

Dans la suite de cette étude de portails internationaux, nous nous limiterons à l'analyse des critères participant de la définition des 4 types de configurations de portail ci-dessus identifiées. La commande DNEUF ayant déjà orienté le choix de solution vers un modèle se rapprochant des portails sup-numerique.gouv.fr et plus particuli7rement les portails UNT, les points de croisement de ceux-ci avec les portails internationaux serait au niveau des quatre points clés suivants :

1. *Modèle d'architecture du portail* : identifier sur lequel des quatre modèles de portails ci-dessus indiqués, le portail international à étudier s'aligne-t-il ;

2. *Procédés de recensement et localisation des ressources* : identifier quelle sources et méthodes de compilation de ressources pédagogiques le portail met-il en place pour constituer sa base de données locale ou distribuée ;

3. *Mécanismes d'indexation* : identifier la technique (et le logiciel) de référencement utilisé par le portail étudié pour indexer les ressources pédagogiques ;

4. *Modes d'accès et de mise à disposition des ressources* : identifier les facilités d'accès mises en œuvre pour accéder et diffuser les ressources pédagogiques.

L'essentiel de cette démarche est de découvrir vers quel type de modèle de portail les institutions choisies (en dehors du contexte français) s'orientent-elles dans la conception de portails de ressources pédagogiques et dans quelles mesures elles présentent des caractéristiques analogues ou différentes à celles proposées par les portails français du sup-numerique.gouv.fr et des UNT.

# PORTAILS INTERNATIONAUX

## 8.3 REFER : RESEAU FRANCOPHONE DE RESSOURCES EDUCATIVES REUTILISABLES

Le projet Réseau francophone de ressources éducatives réutilisables (REFRER) est choisi pour la similitude de son contexte de mise en œuvre avec la commande DNEUF, celle d'élargir une expérience de portail francophone existant au Nord à des partenaires francophones du SUD.

REFRE est un projet soutenu financièrement par l'Organisation internationale de la Francophonie. Mis en œuvre par le Centre de recherche LICEF de la Télé-université (TÉLUQ), il rassemble sept partenaires de quatre pays membres de la Francophonie (Canada-Québec, France, Maroc et Tunisie).

Parmi les avantages technologiques clés du portail REFER notons surtout :

- Offrir des accès à des milliers de ressources éducatives libres par l'utilisation de systèmes de moissonnage performants sur l'Internet ;
- Faire cohabiter l'outil ORI-OAI (France) et l'outil COMÈTE (Québec) de gestion des banques de ressources sur un même portail pédagogique ;
- Valoriser et rendre obligatoire l'utilisation des licences libres (Creative Commons, GNU)
- Stimuler le référencement des ressources selon la norme ISO-MLR ;
- Favoriser le développement d'une infrastructure sur le Web de données liées ;
- Configurer et mettre en place une interconnexion opérationnelle de plusieurs référentiels de ressources éducatives libres.

Dans sa conception globale, REFER se rapproche de la perspective tracée par la commande DNEUF, notamment sur le détail stratégique d'impliquer des partenaires francophones du Sud dans un réseau d'accès et de partages de ressources pédagogiques en ligne. L'étude approfondie de cette expérience est recommandée pour la commande DNEUF.

| Portail | 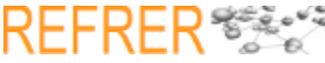 http://www.refrer.licef.ca/ |
|---|---|
| Modèle d'architecture du portail | REFER est un portail constitué de banques de ressources pédagogiques mises en réseaux de serveurs informatiques communiquant via Internet. C'est donc un portail de fédéré entre des établissements disposant |

| | chacun de son propre patrimoine documentaire (ou référentiel) qu'il met en partage via un portail central de recherche et d'accès aux ressources pédagogiques numériques. |
|---|---|
| | Dans son architecture globale, le portail a été conçu sur plusieurs étapes : |
| | 1. Démarrage et mise en place des équipes |
| | 2. Adaptation de la méthode Q4R de production des guides |
| | 3. Analyse de l'environnement technique et intégration d'ORI-OAI |
| | 4. Localisation et évaluation des ressources à intégrer |
| | 5. Installation sur serveur et indexation des ressources |
| | 6. Validation technique du réseau |
| | 7. Enregistrement des banques et mise en réseau |
| | 8. Formation des utilisateurs du réseau |
| | 9. Analyse de deux bancs d'essai des cas d'utilisation du réseau |
| | 10. Complétion du portail de projet pour l'information technologique |
| | 11. version finale de la méthodologie et des instruments pour l'extension du réseau |
| | 12. Organisation de rencontres de diffusion des résultats du projet |
| | |
| Procédés de recensement et localisation des ressources | Une bonne part des ressources du portail existe déjà dans les référentiels TÉLUQ/VTÉ (Canada-Québec) et UNIT (France). Les nouveaux partenaires du SUD (Tunisie et Maroc) ont été associés au projet en vue de développer leurs propres banques de ressources (données et métadonnées) hébergés sur des serveurs dédiés (500 ressources réunis par le Maroc et la Tunisie). Ces ressources ont été créées ou adaptées par des enseignants de ces pays, dans des disciplines déterminées par les partenaires tunisiens et marocains.<br><br>Pour construire le portail commun entre les membres REFER, il fallait mettre en place une infrastructure technologique de réseau de banques de ressources et une méthodologie de mise à disponibilité et de partage des ressources. |
| Mécanismes d'indexation | Une fois les banques de ressources mises en place, un processus d'indexation et de moissonnage de métadonnées entre outils et plateformes du portail a été entamée grâce à un ensemble de normes de métadonnées interopérables (LOM, DC, MLR) et de profils d'application (Normétic, SupLOMfr, Profil Maroc, Profil Tunisie...).<br><br>Le partage des ressources a été mis en œuvre grâce à deux outils de gestion de ressources, ORI-OAI pour la partie française et maghrébine, et COMÈTE pour la partie canadienne.<br><br>Le système ORI-OAI permet d'indexer les ressources selon divers schémas XML de métadonnées notamment le format d'indexation LOM et ses différentes déclinaisons françaises (LOM-FR et SupLOM-FR par exemple). Cette opération permet d'assurer le partage et la diffusion des fiches descriptives obtenues, et d'organiser en aval la recherche au sein |

| | des collections documentaires ainsi constituées. |
| --- | --- |
| | COMÈTE, l'équivalent canadien d'ORI-OAI, s'inscrit dans la mouvance du web sémantique et s'appuie surtout sur la nouvelle norme ISO-MLR (Metadata for learning resources) du SC36. Il adopte l'approche RDF (Resource Description Framework) pour pallier les difficultés posées par la multiplicité des formats, des vocabulaires et des normes de référencement, donc à faciliter le moissonnage et la recherche des ressources. |
| | Le système COMÈTE est bâti pour : |
| | - accepter divers formats de métadonnées (LOM, Dublin Core,...). |
| | - Gérer les identités des personnes et des organisations. |
| | - Gérer des vocabulaires diversifiés |
| | - Permettre des recherches plus ciblées. |
| | - Ouvrir les données au "Linked Data" (données liées). |
| Modes d'accès et de mise à disposition des ressources | Les référentiels sont interconnectés et il possible de faire une recherche unifiée parmi les ressources référencées et versées jusqu'à ce jour dans les six référentiels. |
| | La recherche des ressources s'accomplit sur REFER de différentes façons : par annuaire thématique, par fonds documentaires (par institution), en utilisant un thésaurus, en effectuant une recherche simple ou une recherche avancée (booléenne). |

## 8.4 OPENEDUCATIONEUROPA

En septembre 2013, la Commission européenne a lancé l'initiative «Open Education Europa» dans le cadre du programme Opening Up Education afin d'offrir un accès unique à des ressources éducatives libres (REL) européennes.

L'objectif principal du portail «Open Education Europa» est de fournir un accès à toutes les ressources éducatives libres européennes de haute qualité en différentes langues afin de pouvoir les mettre à disposition des étudiants, des enseignants et des chercheurs. Le but ultime est d'être en mesure de favoriser une création et une utilisation large des REL (Ressources Educatives Libres) en plusieurs langues, pour tous les secteurs et disciplines d'enseignement, et pour aider à surmonter la fragmentation actuelle de l'utilisation des REL en Europe.

« Open Education Europa » est une plateforme dynamique fondée sur les dernières technologies de pointe en matière de licences libres, offrant des instruments pour la communication, le partage et la discussion. Le portail se compose de trois sections principales :

1. La section «Trouver» présente les MOOC, les cours et les ressources éducatives libres des principales institutions européennes. Chaque institution est également présentée dans cette section, tout comme les MOOC, les cours et les REL qu'elle fournit.

2. La section «Partager» est l'espace où les utilisateurs du portail (chercheurs, éducateurs, décideurs politiques, étudiants et autres parties prenantes) peuvent partager et discuter de solutions à divers problèmes éducatifs en postant des blogues, en partageant des évènements, et en participant à des discussions thématiques.

3. La section «En savoir plus» contient les articles d'eLearning Papers – le magazine en ligne sur l'éducation ouverte et les nouvelles technologies le plus visité au monde –, fournit une liste complète des projets financés par l'Union européenne et présente les dernières nouvelles concernant l'éducation ouverte, ainsi que les derniers articles les plus pertinents publiés par des spécialistes.

| Portail | Open**Education**Europa |
|---|---|
| | http://openeducationeuropa.eu |
| Modèle d'architecture du portail | En parcourant les différents services du portail, on peut rapidement détecter les traces d'une organisation fédérative autour d'un ensemble de serveurs de données qui, en s'inscrivant au projet du portail OpenEducationEuropa, donnent accès à des ressources variées : Moocs, cours, ressources, données institutionnelles, blogs, discussions, revues, évènements etc. |
| | *(Diagramme OpenEducationEuropa : Institutions / Alliances — EDEN, Study Portals, EADTU (access to all Open Universities), Open CourseWare Consortium Europe (access to all European HE Institutions that follow OCW model) ; Over 200 HE Institutions in Europe (28+ countries) ; ANY OTHER INSTITUTIONS ; MOOC Platforms / OER Repositories / Other Initiatives ; Miríada X, Un X, EdX (European MOOCs), Open Course World, OpenUpEd, FutureLearn, Coursera (European MOOCs), Iversity)* |
| Procédés de recensement et localisation des ressources | Ce type de portail procède par une politique de partenariat avec des serveurs tiers qui décident du type de ressources libres à rendre accessibles ou à soumettre au moissonnage par le moteur du portail.<br><br>La majorité des réservoirs des ressources européennes offrent des volumes de REL libres qui peuvent aller de 100 à plus de 300 000 REL publiées. La plus grande majorité des ressources sont fournis sous la licence Creative commons. |
| Mécanismes d'indexation | Toutes les URL des ressources consultées conduisent à des serveurs externes. C'est l'un des indicateurs que le portail ne dispose pas d'un dépôt de ressources mais plutôt d'un référentiel de ressources (réservoir de métadonnées).<br><br>L'autre indicateur est qu'avant d'accéder à une ressource externe, le portail affiche sa notice descriptive (métadonnées) avec un lien vers la source externe. Il est fort probable (aucune documentation technique ne le signale) que ce référentiel soit le résultat d'un protocole de moissonnage du type OAI-PMH.<br><br>La normalisation n'est pas encore à l'ordre du jour (taxonomies, listes de vedettes matières). |
| Modes d'accès et de | Le portail dispose par contre de son propre système d'information très |

| mise à disposition des ressources | développé, riche en services de référencement/recherche et en fonctions à valeurs ajoutées. |
|---|---|
| | Il dispose d'un puissant moteur de recherche avec des filtres intelligents afin que les utilisateurs trouvent facilement ce qu'ils recherchent : des REL, cours et Moocs structurés. |
| | Le portail offre une interface multilingue et un contenu multilingue créé en collaboration avec les meilleures institutions en Europe. Le contenu concerne tous les niveaux d'enseignement. |

## 8.5 AUTRES PORTAILS : REMARQUES GENERALES

Plusieurs autres portails de ressources pédagogiques ont été visités pour repérer si des formes radicalement différentes des exemples étudiés jusqu'ici peuvent exister. En voici une liste brève :

- ✓ **EDUSCOL : Portail national des professionnels de l'éducation :** Educscol héberge plusieurs systèmes de gestion de ressources pédagogiques

- ✓ **PRIMTICE : Portail des TICE pour l'école primaire (**http://primtice.education.fr/) site du ministère de l'éducation nationale pour accompagner les usages du numérique dans le premier degré. PrimTICE s'appuie sur un répertoire de plusieurs centaines de scénarios pédagogiques développés par les enseignants et mettant en œuvre les TICE, de la maternelle au cycle 3. Cette base, accessible en ligne, sert de supports aux enseignants qui conçoivent des projets, ou des séances faisant appel aux nouvelles technologies.

- ✓ **CEL – CCSD Cours en ligne (**https://cel.archives-ouvertes.fr/) destiné à offrir aux doctorants qui travaillent dans les laboratoires l'accès à des cours qui peuvent leur être utiles : cours de DEA et master-recherche, des grandes écoles, écoles d'été par exemple. La consultation est libre et gratuite.

- ✓ **CANOPE (**http://www.reseau-canope.fr/) Réseau de création et d'accompagnement pédagogiques est un éditeur de ressources pédagogiques public, dépendant du ministère de l'Éducation nationale français. Il succède depuis 2014 au réseau SCEREN (Services, culture, éditions, ressources pour l'Éducation nationale).

- ✓ **REL@UVA : Ressources éducatives libres de l'Université Virtuelle Africaine (**http://oer.avu.org/) une initiative de l'Université Virtuelle Africaine qui prend en charge les questions touchant à la pertinence des REL pour le contexte africain.

- ✓ **MIT OPEN COURSEWARE (**http://ocw.mit.edu/index.htm) est une publication sur le Web de pratiquement tous les supports de cours du MIT. OCW est ouvert et disponible à tous. Il constitue une activité permanente de MIT.

- ✓ **Open Courseware Consortium (**http://www.oeconsortium.org/) offre à ses membres les outils et ressources pour développer leur propre contenu.

- ✓ **MERLOT: teacher educational portal (**http://teachereducation.merlot.org/) un consortium international de plus de 20 institutions (et systèmes) de l'enseignement supérieur, de partenaires de l'industrie, d'organisations professionnelles et d'individus consacrés à l'identification, l'évaluation par les pairs, l'organisation et la mise à disposition de ressources WWW existants dans un éventail de disciplines universitaires.

- ✓ **TEMOA: open educational resource portal** est un centre de connaissances qui facilite l'accès à un catalogue public et multilingue de ressources éducatives libres (REL). Il contient une sélection des ressources éducatives, décrites et évaluées par une communauté académique. Il fournit un moteur de recherche convivial à travers des filtres intuitifs.

- ✓ **OER COMMONS : Open Education Resources (**https://www.oercommons.org/) une bibliothèque en ligne accessible gratuitement qui permet aux enseignants de rechercher et de découvrir des ressources éducatives ouvertes (OER) et autres matériels pédagogiques disponibles gratuitement.

L'ensemble de ces portails et de services d'accès et de diffusion de ressources pédagogiques libres reproduisent (avec des variations de nature plutôt ergonomique que structurelle) les modèles de portails identifiés au moment de la synthèse des portails sup-numerique.gouv.fr et UNT (portails hybrides, de références, vitrines ou autonomes). La tendance est plutôt vers les portails de référence, moins couteuses en production et en entretien que les portails autonomes, mais plus efficace et crédibles que les portails vitrines démunis de toutes ressources et de toutes références.

Le choix d'une structure particulière pour un portail de ressources pédagogiques dépendra d'un ensemble de critères que seules une étude de faisabilité et un cahier des charges permettraient de définir la viabilité économique, l'efficacité technique et la rentabilité scientifique.

A l'issu de cette étape, en en tenant compte de la commande DNEUF qui pose les condition d'orientation donnée à cette étude, la section suivante sera réservée à une série de recommandations pour le groupe d'experts de l'AUF chargé d'établir le cahier des charges d'un portail francophone commun et de proposer les modes possibles de sa réalisation.

# 9 RECOMMANDATIONS

Les recommandations suivantes concernent le cadre d'organisation et les mesures technologiques à entreprendre pour la conception et la réalisation du système d'information du portail et non de son design graphique et encore moins de son optimisation ergonomique. Ces dernières mesures feraient l'objet d'une phase ultérieure quand le choix d'une structure définitive au portail serait pris et validé.

Ces recommandations sont regroupées en quatre ensembles de points concernant le modèle d'architecture envisagé pour le portail francophone, les procédés de recensement et de collecte des ressources pédagogiques, les mécanismes de référencement et d'indexation de ces ressources puis le mode d'accès et de mise à disposition de ces ressources en ligne. Une série de recommandations à caractère général (de gestion et d'administration de portail) viendrait boucler ce premier rapport d'analyse de l'existant en portails de ressources pédagogiques libres.

| Modèle d'architecture du portail | - L'étude des portails sup-numerique.gouv.fr, UNT et autres portails internationaux montrent qu'il y a quatre types courant d'organisation de portail de ressources pédagogiques : hybrides, autonomes, vitrines et de références. Le choix de l'un ou de l'autre dépend d'un ensemble de critères institutionnels, économiques, politique et humains qu'il faudrait définir en amont dans un cahier des charges. Par exemple, si le portail IUTenligne est exceptionnellement autonome, c'est entre autres, parce que son cadre institutionnel l'exige. Les IUT constituent un réseau d'institutions nationales plus homogènes qui travaillent en communauté quasi fermée et juridiquement mieux ancrée qu'un ensemble d'universités liées par des accords ou des conventions changeantes. Ils ont un public cible spécialisé d'étudiants et d'enseignants tenus par les mêmes types de besoins et de procédures de travail. Ils produisent et exploitent leurs propres ressources pédagogiques dans des conditions d'enseignement et de recherche très apparentées ;<br><br>- Dans le même ordre d'idées, si le portail sup-numerique.gouv.fr s'est voulu une vitrine universitaire transversale et hybride, c'est en raison de deux éléments clés. D'abord, la vocation initiale de sup-numerique.gouv.fr est d'exploiter les ressources pédagogiques déjà référencée par les portails des UNT (donc utiliser le référentiel ou métadonnées des ressources existantes des universités membres). Ensuite, la deuxième vocation de sup-numerique.gouv.fr est de créer ses propres dépôts de ressources (les cours Moocs) et de les intégrer dans un référentiel général (métadonnées) qui englobe ses propres ressources et celles des UNT. Son nouveau moteur de recherche a été conçu pour s'acquitter de cette charge grâce aux techniques de référencement ORI-OAI ;<br><br>- Le portail francophone de ressources pédagogiques devrait donc observer la diversité des partenaires associés et du public hétérogène à servir et la traduire sous forme de services et de fonctions qui conviennent à tous les publics concernés ;<br><br>- De ce fait, avant de choisir une configuration particulière dans le cadre de la commande DNEUF, l'AUF et l'IFIC sont tenus de faire une étude de l'existant (dont ce rapport constitue un prélude). Il est |
|---|---|

nécessaire dans un stade ultérieur, et avant d'opter pour une configuration donnée, de dresser un inventaire préliminaire des acteurs institutionnels concernés, des ressources humaines nécessaires, des moyens financier et techniques à pourvoir, des partenaires académiques et politiques à engager, de la nature des services à développer et de la durée de réalisation de l'ouvrage, des type de ressources pédagogiques à retenir pour le projet et des populations cibles à servir, etc. Cette étude aidera à tracer les grandes lignes d'un scénario de conception/réalisation d'une architecture de portail de ressources pédagogiques francophones ;

- Eu égard à la nature des activités francophones en enseignement à distance, à la nature de ses partenariats internationaux et à l'envergure de son champ d'action et d'intervention dans les stratégies éducatives francophones et internationales, l'AUF et l'IFIC gagneraient à adopter une configuration hybride pour deux raisons simples :

    1. d'abord, il y a un patrimoine pédagogique très riche mais éparpillé et mal exploité dans les institutions universitaires francophones. Ce patrimoine doit être récupéré par des opérations d'assainissement méthodiques en les soumettant à des opérations *in situ* de recensement, traitement, mise à jour et mise en forme cohérente pour les intégrer dans une construction interopérable par les normes technologiques en vigueur ;

    2. ensuite, l'AUF & l'IFIC sont producteurs ou coproducteurs de ressources pédagogiques numériques (projets de recherche, ateliers de formations, politique de numérisation, etc.) qu'ils devraient mettre à la disposition de la communauté francophone. Ils devraient aussi fournir à leurs partenaires francophones, notamment du Sud, un accès à la production pédagogique internationale non francophone. Un moteur de recherche sur le modèle sup-numerique.gouv.fr serait très approprié sous réserve que les institutions partenaires mettent à niveau leurs propres dispositifs (i.e. harmonisation des métadonnées, installation d'un moteur local ORI-OAI) pour se conformer à des règles de bonnes pratiques collaboratives et à des techniques normatives de systèmes d'information interopérables.

| | |
|---|---|
| Procédés de recensement et de | - Les ressources à intégrer au portail doivent être identifiées par un référentiel (inventaire) typologique et quantitatif (genre et volume |

| sélection de ressources | de chaque type de ressource) qui doit remonter depuis chaque institution partenaire vers l'équipe de pilotage du projet. Les ressources peuvent prendre des formes multiples. Le référentiel pourrait être spécifique à un ou plusieurs types de ressources (i.e. cours, Quizz, scénarios pédagogiques), une ou plusieurs collections, un ou plusieurs domaines ou simplement tous les types de produits éducatifs. Le choix des ressources devrait s'effectuer normalement selon une politique éditoriale proposant des critères de sélection précis et partagés par les partenaires concernés. Ces critères concerneraient les dimensions pédagogiques, culturelles, techniques et ergonomiques qui assureraient la qualité des ressources intégrées ; |
|---|---|

- L'AUF et l'IFIC doivent prévoir la mise en œuvre d'une convention de travail avec les institutions universitaires francophones pour l'harmonisation des procédés de travail documentaire des fonds et collections des ressources pédagogiques locales. Cette convention définirait des règles de bonnes pratiques fondées sur des normes et des standards internationaux. Parmi ces normes et standards, l'application d'un profil d'application francophone de métadonnées pédagogiques qu'il faudrait définir d'urgence (cf. sous Mécanisme d'indexation) ;

- Une étude exploratoire devrait aussi être menée à l'échelle internationale pour identifier des portails pédagogiques existant qui produiraient et fourniraient un accès libre à une matière pédagogique complémentaire jugée utile. Des accords et des conventions avec ces portails devraient être proposés pour les intégrer dans le champ de couverture du portail des ressources pédagogiques francophone à construire. Le système d'information du nouveau portail devrait pouvoir encapsuler plusieurs systèmes d'information hétérogène (en plus du sien) sous un même dispositif d'accès libre à des ressources à la fois locales et dispersées, sans tenir compte de leurs environnements techniques. En associant ORI-OAI et COMETE dans un même dispositif, le projet REFER a montré la faisabilité d'une pareille procédure technique. Le schéma de fonctionnement de REFER fait apparaître un accès au réseau GLOBE, qui regroupe plusieurs portails fédérateurs de plusieurs banques de ressources, tels qu'ARIADNE (Europe), MERLOT (USA) ou LACLO (Amérique latine). Les outils gestionnaires de banques de ressources sur REFER offrent des services informatiques rendant possible une recherche fédérée ou un moissonnage des banques de ressources sur Internet, comme s'il s'agissait d'une seule banque. Ainsi, par exemple, un enseignant peut sélectionner et afficher les ressources qui l'intéressent, y accéder, les adapter à ses besoins et les regrouper pour créer des cours, des pages Web, des sites d'apprentissage pour ses étudiants, des webographies. Ces scénarios de cours sont eux-

| | |
|---|---|
| | mêmes des ressources pouvant être à leur tour intégrées dans une banque de ressources en décrivant leurs métadonnées ;<br><br>- La typologie des ressources à concevoir et à intégrer au portail est aussi sujette à décision, car elle pose des questions multiples afférant aux niveaux scolaires et domaines de formation, à la qualité et validation des contenus scientifique, à la granularité sémantique et technique des ressources, aux langues des supports, etc. Une typologie bien identifiée des ressources et des conditions et modalités de leur intégration dans le futur portail devrait faire l'objet d'un rapport circonstancié. Ce rapport tiendrait compte de besoins des établissements francophones partenaires et des exigences de l'AUF/IFIC dans leurs stratégies de politique éducatives. |
| Mécanismes d'indexation | - Deux points clés sous cet angle sont à déterminer dans la prise de décision de la part de l'AUF/IFIC concernant le portail francophone des ressources pédagogiques : le protocole général d'indexation à choisir pour faciliter la recherche et la récupération des ressources puis le modèle de référencement interopérable à adopter conformément à une norme internationale de métadonnées pédagogiques ;<br><br>- Le choix du protocole général d'indexation est quasi déterminé par la commande DNEUF qui veut établir une extension du modèle sup-numerique.gouv.fr et UNT au projet de portail francophone. La solution ORI-OAI conforme OAI-MPH (au cœur des portails sup-numerique.gouv.fr et UNT) semble faire l'anonymat en France, en Europe et dans certains pays francophones d'Afrique, contrairement à d'autres solutions implantées aux USA et au Canada (COMETE, EUREKA, PALOMA, FEDORA). Par souci de conformité à un choix environnant (Franc, Europe), le choix d'ORI-OAI est à double avantage : d'une part gagner l'aval et l'appui des institutions et des communautés opérant déjà sous ORI-OAI, et d'autre part, rester ouvert aux, et être interopérable avec, toutes autres institutions et communautés opérant sous d'autres systèmes de référencement conformes à OAI-PMH ;<br><br>- Il est à rappeler dans ce sens qu'ORI-OAI propose une implémentation du protocole OAI-PMH et d'outils nécessaires à la création de nouvelles communautés et à la bonne intégration aux communautés existantes ;<br><br>- Il est à rappeler aussi que les objectifs de l'intégration d'ORI-OAI dans un établissement peuvent être différents selon que l'on se sert de cet outil dans la perspective d'un portail de ressources numériques ou dans celle d'une archive institutionnelle. Dans le cas d'un portail de ressources pédagogiques en ligne, ORI-OAI permet à |

- la fois la publication des ressources produites par l'établissement concerné et la diffusion et le partage de ces ressources avec d'autres établissements et institutions, grâce notamment au protocole OAI-PMH. C'est là l'essentiel du rôle défini pour le portail francophone de ressources pédagogiques comme envisagé par la commande DNEUF ;

- Par rapport au deuxième point (le modèle de référencement interopérable selon une norme internationale de métadonnées pédagogiques), la force d'ORI-OAI est de permettre d'indexer des ressources selon divers schémas XML de métadonnées, puis d'assurer le partage et la diffusion des fiches descriptives obtenues, et enfin, d'organiser la recherche au sein des collections documentaires ainsi constituées ;

- L'inconvénient d'ORI-OAI, par contre, est que malgré son adaptation à couvrir tous les types de ressources numériques et à supporter le format d'indexation LOM et ses différentes déclinaisons (par exemple LOM-FR et SupLOM-FR en France, CANCORE et NORMETIC au Canada)[9], aucune expérience n'a encore démontré sa capacité à supporter un profil d'application issu de la nouvelle norme MLR. Des produits comme COMÈTE[10], Eurêka[11] et Paloma[12] sont, quant à eux, fondés sur la norme ISO/IEC SC36 19788 « METADATA FOR LEARNING RESOURCES (MLR) ». Ils sont adaptés au stockage des métadonnées sous la forme de triplets RDF (RESOURCE DESCRIPTION FRAMEWORK).Ils marquent par cela un point d'avance par rapport à ORI-OAI ;

- Il faudrait rappeler en effet, qu'en 2011, une nouvelle norme internationale de métadonnées pédagogiques a été publiée par l'ISO (International Standards Organisation) : « MLR - *Metadata for Learning Resources »* (Métadonnées pour ressources d'apprentissage - ISO/IEC 19788). La norme MLR repose sur le standard RDF - *Resource Description Framework*, développé par le W3C pour faciliter le traitement des métadonnées. MLR vise à intégrer les

---

[9] Ben Henda Mokhtar (2010). « Pour un Programme d'Appui à l'Interopérabilité Universitaire en Tunisie : rôle des normes et des standards d'interopérabilité pour les technologies éducatives et l'e Learning ». HAL Archives ouvertes, https://hal.archives-ouvertes.fr/sic_00523345/document

[10] Comète est une application logicielle permettant de trouver, agréger, organiser en collection et diffuser le patrimoine numérique des institutions d'enseignement et autres intervenants des systèmes d'éducation, http://comete.licef.ca/Portal/?lang=fr

[11] Le projet Eurêka est une initiative dans le cadre d'un projet de coopération Québec-Wallonie-Bruxelles qui offre un catalogue collectif de ressources d'enseignement et d'apprentissage rassemblées par divers organismes œuvrant dans la production de ressources éducatives TIC. http://www.eureka.ntic.org/

[12] PALOMA est un gestionnaire de métadonnées. Il permet de référencer les objets d'apprentissage selon les métadonnées du LOM, de SCORM, de CANCORE et de Normetic. De plus, ce système a la capacité de se relier à différentes banques de métadonnées et d'éditer les métadonnées sur lesquelles il a la permission. http://helios.licef.ca/PalomaSuite/index.htm

| | standards Dublin Core et LOM dans la perspective ouverte du Web de données (LINKED DATA) ou Web sémantique. Cette initiative constitue un pas géant vers les réseaux sémantiques qui sont au cœur de la société de la connaissance et du savoir partagé, l'un des axes prospectifs sur lequel l'AUF construit sa stratégie d'avenir ; |
|---|---|
| | - Il faudrait rappeler aussi que l'idée du passage à un profil d'application MLR est ancienne. En 2004, Yolaine Bourda, coéditrice de MLR au SC36, annonçait déjà que « *L'enjeu consiste à faire en sorte que le MLR soit compatible avec le LOM – afin de préserver ce qui est déjà acquis – tout en prenant en compte certains points qui ne sont pas abordés dans le LOM… Dans le LOM, le nom d'un champ est confondu avec sa signification, or celle-ci n'est pas neutre : elle dépend du contexte, de la culture, de la langue. L'idée est donc de s'intéresser aux définitions à un niveau abstrait, afin que tous, quels que soient leur culture et leur langue, puissent s'accorder sur la signification de chaque élément, sans se focaliser sur son nom. Une fois que la signification d'une valeur est bien définie, il est possible de lui donner plusieurs noms, ce qui résoudrait le problème du multilinguisme* »[13] ; |
| | - Le point précédent conduit inéluctablement à l'urgence de procéder au développement d'un profil d'application francophone auquel se conformeraient (ou du moins sur lequel s'aligneraient) toutes les institutions universitaires et de recherche francophones productrices de ressources pédagogiques numériques. L'AUF et l'IFIC devraient donner une haute priorité à cette démarche pour produire un outil stratégique d'harmonisation, de convergence et d'interopérabilité entre les institutions qui œuvrent en synergie avec l'AUF et la Francophonie en général (i.e. OIF). La démarche qualité du futur portail de ressources pédagogiques en dépendrait profondément. |
| Modes d'accès et de mise à disposition des ressources | - Plusieurs modes d'accès et de diffusion des ressources pédagogiques libres peuvent être envisagés : au grand public sans aucune restriction identitaire ou de profil (modèle des archives ouvertes) ou sur justificatif d'affiliation à une institution ou un programme de formation (*a minima* via une inscription publique et *a maxima* via un LDAP d'un ENT institutionnel) ; |
| | - Respectant le principe des REL (ressources éducatives libres), l'accès gratuit et ouvert aux ressources pédagogiques en ligne devrait être une constante politique de la Francophonie. Cependant, dans le cas d'un portail hybride (ressources propres et ressources liées) les institutions partenaires peuvent imposer leurs politiques d'accès |

---

[13] Bourda, Yolaine. « L'indexation des ressources pédagogiques », 2004, enssib, Villeurbanne. http://www.enssib.fr/bibliotheque-numerique/documents/1240-les-evolutions-du-lom.pdf

locales. Le portail sup-numerique.gouv.fr en donne l'exemple quand il permet un accès ouvert et gratuit à ses propres ressources Moocs (même si une inscription est nécessaire pour accéder aux cours d'un Mooc) mais quand il s'agit de donner accès aux ressources des UNT, le portail sup-numerique.gouv.fr applique la politique d'accès propre à ses partenaires. Alors que la majorité des portails UNT sont d'un accès ouvert et gratuit aux métadonnées et aux ressources, celui de l'Université numérique juridique francophone (UNJF) donne accès uniquement aux métadonnées et réserve l'accès aux ressources aux inscrits dans les annuaires des instituions partenaires[14].

- Le portail francophone de ressources pédagogiques libres devrait disposer de son propre système d'information assez développé et riche en services de référencement/recherche et en fonctions à valeurs ajoutées parmi lesquelles :

    ✓ Des outils de traduction (des résultats de recherche et des contenus) ;

    ✓ Des choix linguistiques pour l'interface de navigation ;

    ✓ Une personnalisation des espaces de l'interface (portlets et widgets) ;

    ✓ Des fonctionnalités d'imports/exports des références en formats XML normalisés (DC, LOM, MLR, Unimarc, Z3950, etc.) ;

    ✓ Des services d'exports de résultats de recherche en formats bibliographiques standards (Zotero, EndNotes, BibText, etc.) ;

    ✓ Des fils RSS et des outils d'alerte ;

    ✓ Des outils de syndication et de commentaires par les usagers (votes, blogs, news, réseaux sociaux, etc.) ;

    ✓ Des filtres intelligents de recherche par autant que possible de critères de recherche combinée (par thèmes, auteurs, dates, langues, institution, types, formats, etc.) ;

- Dans une démarche innovante de diffusion et de mise à disponibilité des ressources pédagogique, le portail francophone gagnerait à mettre en œuvre une fonctionnalité importante qui permet (via des API spécifiques) de lier une ressource localisée sur le portail aux flux des données générés par les dispositifs et plates-formes pédagogiques et les Environnements numérique de travail (ENT) les plus utilisés dans le contexte francophone, notamment Moodle,

---

[14] UNJF donne exceptionnellement un accès gratuit à certaines ressources aux visiteurs invités au moyen d'un nom d'utilisateur et d'un mot de passe attribués sur demande en ligne.

Cocktail, etc. Cette fonction gagnerait à se faire selon le modèle d'agrégation de la norme SCORM (ISO/IEC TR 29163:2009).

| | |
|---|---|
| Mesure générales (gestion et administration de portail) associées aux points précédents | Pour un portail francophone de ressources pédagogiques en accès libre, l'AUF et l'IFIC devraient observer les mesures suivantes :<br><br>- Autour du modèle d'architecture du portail :<br><br>  ✓ La commande DNEUF part du prototype sup-numerique.gouv.fr /UNT pour élargir ses fonctionnalités au futur portail francophone à mette en place. Cette décision de devrait pas ignorer l'étude d'autres solutions internationales qui pourraient enrichir davantage le prototype de portail à concevoir. Une synthèse de certaines fonctionnalités non proposées par sup-numerique.gouv.fr/UNT a été faite dans ce rapport. D'autres fonctions pourraient faire l'objet d'une attention plus concentrée au moment de la rédaction du cahier des charges du portail francophone ;<br><br>  ✓ Il est nécessaire d'envisager une équipe permanente de suivi du portail qui impliquerait a minima un administrateur réseau, un documentaliste, un spécialiste e-learning et de pédagogie universitaire ;<br><br>- Autour des procédés de recensement et de sélection de ressources<br><br>  ✓ Former une équipe dans chaque institution pour assurer des rôles de gestionnaire, d'évaluateur, de contributeur/auteur et ou d'indexeur ;<br><br>  ✓ Elaborer un manuel de procédures au profit des partenaires universitaires pour rappeler les directives fondamentales concernant la production harmoniée des ressources et leur traitement comme des ressources interopérables et partageables en ligne ;<br><br>  ✓ Mettre en place des moyens incitant et valorisant les fournisseurs de ressources (auteurs individuels et institutions) ;<br><br>  ✓ Valoriser et rendre obligatoire l'utilisation des licences libres (CreativeCommons, GNU).<br><br>- Autour des mécanismes d'indexation<br><br>  ✓ Favoriser le référencement intégré des ressources selon les normes internationales (DC, LOM, MLR, LRMI[15]) |

---

[15] L'Initiative de métadonnées des ressources d'apprentissage (LRMI) est un projet mené par Creative Commons (CC) et l'Association of Educational Publishers (AEP) et soutenu par plusieurs acteurs importants de l'industrie

- ✓ Créer une équipe de travail (*Task force*) chargée de proposer et développer un profil d'application MLR francophone ;

- ✓ Développer le plus largement possible un vocabulaire commun pour le besoin des opérations d'indexation/recherche des ressources pédagogiques ;

- ✓ Travailler sur le long terme sur la production d'une ontologie du domaine des technologies éducatives et du e-learning pour favoriser l'intégration des REL au Web de données liées ;

- Autour des modalités d'accès et de mise à disposition des ressources

  - ✓ Installer un outil (ORI-OAI) au niveau du serveur des données de chaque institution partenaire et l'interfacer à un environnement numérique de travail ;

  - ✓ Rendre disponible des aides méthodologiques et de la formation au profit des institutions partenaires pour la mise en place de l'outil ORI-OAI et l'interconnexion des référentiels de ressources éducatives libres ;

- Recommandation générale

  - ✓ La réalisation du portail devrait se faire sur la base de conventions, accords et règlements de partenariats solides qui définissent clairement les obligations et les droits de chacun des acteurs engagé dans la conception, réalisation, administration et usage du portail. Il importe de définir clairement les politiques que les parties prenantes s'engageront à respecter, d'obtenir un commun accord et une compréhension commune sur la propriété intellectuelle et sur la responsabilité de chacun, ainsi que sur l'édition et la contribution des ressources. Les politiques sont importantes pour clarifier les responsabilités des acteurs, faciliter la planification des opérations et choisir/construire les outils d'assistance. Cette politique doit expliquer comment établir une procédure suivant laquelle chaque ressource est évaluée avant d'être acceptée par le référentiel. Les politiques du référentiel précisent en général les exigences pour qu'une ressource soit acceptée.

---

informatique dont Microsoft et Google, pour établir un vocabulaire commun pour décrire les ressources d'apprentissage.

# 10 REFERENCES

Ce rapport a été conçu à l'aide des sites et des lectures de références suivantes qui ont permis de faire le tour de la question des modèles de fonctionnement des portails de ressources pédagogiques et des méthodes de leurs conception.

- Portail UVED de l'environnement et du développement durable http://univ-numerique.fr/developpement-durable/
- Pour le développement numérique de l'espace universitaire francophone : http://www.francophonie.org/Pour-le-developpement-numerique-de.html
- Projet de déclaration commune des ministres francophones : http://www.francophonie.org/IMG/pdf/declaration_ministres_enseignement_ressources_numeriques.pdf
- Projet Eurêka. http://www.eureka.ntic.org/
- site EducNet : http://www.educnet.education.fr/dossier/metadata/lom1.htm